\documentclass[a4paper,11pt]{article}
\pdfoutput=1
\usepackage{jcappub}
\usepackage{amsmath}
\usepackage{float}
\usepackage{graphicx}
\usepackage{epsfig}
\usepackage{latexsym, amssymb} 
\usepackage{amsfonts}
\usepackage{amssymb}
\usepackage{hyperref}
\usepackage{comment}

\def\mpl{M_{\rm Pl}}
\newcommand{\bea}{\begin{eqnarray}}
\newcommand{\eea}{\end{eqnarray}}
\newcommand{\be}{\begin{equation}}
\newcommand{\ee}{\end{equation}}
\newcommand{\beq}{\begin{equation}}
\newcommand{\eeq}{\end{equation}}

\begin{document}

\title{Revisiting CMB constraints on Warm Inflation}%

\author[a]{Richa Arya,}
\author[a]{Arnab Dasgupta,}
\author[b]{Gaurav Goswami,}
\author[c]{Jayanti Prasad,}
\author[a]{Raghavan Rangarajan}

\affiliation[a]{Theoretical Physics Division, Physical Research Laboratory, Ahmedabad 380009, India}
\affiliation[b]{School of Engineering and Applied Science, Ahmedabad University, Ahmedabad 380009, India}
\affiliation[c]{Inter-University Centre for Astronomy and Astrophysics, Pune 411007, India}

\emailAdd{richaarya@prl.res.in, arnabd@prl.res.in, gaurav.goswami@ahduni.edu.in, jayanti@iucaa.ernet.in, raghavan@prl.res.in}

\date{today}

\abstract{We revisit the constraints that Planck 2015 temperature, polarization and lensing data   
impose on the parameters of warm inflation.  To this end, we study warm 
inflation driven by a single scalar field with a quartic self interaction potential in the weak dissipative regime.
We analyse the effect of the parameters of warm inflation,
namely, the inflaton self coupling $\lambda$ and the inflaton dissipation parameter $Q_P$ 
on the CMB angular power spectrum.
We constrain 
$\lambda$ and $Q_P$ 
for 50 and 60 number of e-foldings with the full Planck 2015 data (TT, TE, EE + lowP  and lensing) 
by performing a Markov-Chain 
Monte Carlo analysis using the publicly available code {\tt CosmoMC} and 
obtain the joint as well as marginalized distributions of those parameters.
We present our results in the form of mean and 
68 \% confidence 
limits on the parameters and also highlight 
the degeneracy between  $\lambda$ and $Q_P$ in our analysis. 
From this analysis we show how warm inflation parameters can be well constrained using the Planck 2015 data.
}

\maketitle

\section{Introduction}

Cosmic inflation 
\cite{Starobinsky:1980te,Kazanas:1980tx,Guth:1980zm,Linde:1981mu,Albrecht:1982wi,Starobinsky:1979ty,Hawking:1982cz,Starobinsky:1982ee,Guth:1982ec}
is a phase of accelerated expansion in the early history of the Universe during which the energy density of the Universe is dominated by the density of a slowly rolling scalar field, the inflaton. 
In the simplest models of this inflationary paradigm, it is assumed that the number density of all the particles becomes negligible during inflation due to which slow roll inflation occurs in an almost perfect vacuum state.
The inflaton field is coupled to other fields so that it may dissipate its energy after inflation and reheat the universe. However one may consider the possibility that dissipative effects are important not only after but also \emph{during}
the slow-roll phase of inflation.  
In that case, under appropriate conditions, there is a thermal bath present during the accelerated expansion also, and 
if the temperature of the bath is greater than the Hubble parameter $H$, then this scenario is known as \emph{Warm Inflation} 
\cite{Berera:1995wh,Berera:1995ie,Berera:1999ws}(see Ref. \cite{LopezNacir:2011kk} for a more recent discussion on dissipation during inflation).

Why should one consider warm inflation?
Firstly, it is natural to consider inflaton couplings not just during
reheating but also during inflation. Furthermore, for some warm inflation models, the
eta problem, namely, the presence of large quantum
corrections to the inflaton potential that ruins its flatness, is 
resolved.  Also, some potentials that are excluded by
Cosmic Microwave Background (CMB) data in the cold inflation scenario are allowed in the
warm inflation scenario (though the converse is also true).
Models of warm inflation usually require
a very large number of fields ($10^{4}-10^{6}$) to be coupled to the inflaton to maintain a thermal
bath with $T>H$ \cite{Berera:1998px,Bastero-Gil:2015nja}, which may be
unattractive.  However recently proposed warm inflation models with the inflaton as a pseudo-Nambu-Goldstone boson require only two additional fields \cite{Mishra:2011vh,Bastero-Gil:2016qru}.

Given the importance of warm inflation, it is then
important to constrain the parameters of warm inflation 
scenarios using Cosmic Microwave Background (CMB) data just like it is done for various cold inflation scenarios. 
In this work, we constrain the parameters of warm inflation driven by a quartic potential using complete Planck 2015 data (temperature, polarization and lensing).
In particular, we focus on obtaining the marginal and joint probability distributions of the parameters involved in the model  
by carrying out a full Markov-Chain Monte Carlo analysis using {\tt CosmoMC}  \cite{cosmomc}, a publicly 
available code.

A quartic  warm inflation model is described by the inflaton's self coupling $\lambda$ and its dissipation rate to other fields described by the dissipation coefficient $\Upsilon$. In the literature there are usually two forms for this dissipation coefficient, $C_T T$ and $C_\phi T^3/\phi^2$.
Here we consider only the latter form, which is obtained from
inflaton decay to lighter particles mediated by heavier particles ($T>m$) 
\cite{Moss:2006gt,BasteroGil:2010pb,BasteroGil:2012cm}
and $C_{\phi}$ is related to the inflaton couplings and the multiplicities of other fields in the model.
More generally,
$\Upsilon = C T^c \phi^{2a}/M_X^{2b}$
with $c+2a-2b=1$ \cite{Ramos:2013nsa}, where $M_X$ is the mass of the intermediate
particle. 
For a model builder, knowing constraints from the data
on $\lambda$ and $C_\phi$ will be very useful.

The power spectrum for quartic warm inflation depends on
$\lambda, C_\phi$ and $\phi_k$,
where the subscript $k$ indicates that quantities are 
evaluated at the time when the  scale $k$ crosses the 
horizon during inflation.  In the literature on warm inflation, the authors have introduced the variable
$Q=\Upsilon/(3H)$. The power spectrum for quartic chaotic
warm inflation then effectively depends on $\lambda, 
Q_k$ and $N_k$, the number of e-foldings of inflation from the end of inflation
to the epoch when the  scale $k$ crosses the horizon.
[Recall that in cold inflation the power spectrum can be made to be 
effectively dependent on $\lambda$
and $N_k$ 
(see, for example, Eq. (8.71) 
of Ref. \cite{Kolb:1990vq} for quartic chaotic inflation).]

For cold inflation, the normalisation of the power spectrum at the
pivot scale $k_P$  fixes the value of $\lambda$, for a fixed $N_P$,
where $P$ indicates that quantities are 
evaluated at the time when the pivot scale $k_P$ crosses the 
horizon during inflation. 
We find that in warm inflation too fixing the normalisation at the pivot scale 
fixes the value of $\lambda$ for fixed $N_P$ and a particular value of $Q_P$.  
We 
find that
for the scales of cosmological interest the power spectrum is close to
power law, and we study the dependence of the primordial  power spectrum on the parameters of the model.  
We then study
the angular power spectrum using the publicly available code {\tt CAMB} \cite{camb}
and 
finally constrain the parameters of the quartic warm inflation model
using {\tt CosmoMC}.
In our analysis, we fix  $N_P$ to be either 50 or 60.

We begin by summarizing the known results about the primordial power spectrum in warm inflation.  
We then express the primordial power spectrum for
the warm inflation model with a quartic potential
as a function of $\lambda, C_\phi$ and $Q_P$,
and then eventually in terms of $\lambda, Q_P$ and $N_P$.  We discuss features of the power spectrum
and correlations between parameters of the model.
Thereafter we present our Monte-Carlo Markov Chain analysis using {\tt CosmoMC}
for which we use the full Planck 2015 data (high-l TT,TE, TE + low-l polarization + lensing)
for our analysis. 
We explain how we constrain the parameters of quartic warm inflation and finally show the obtained joint and marginal probability distributions of $\lambda$ and $Q_P$.  We provide the mean and standard deviation for these parameters and highlight the degeneracy between the parameters.

\section{Evolution equations in warm inflation}

For the inflaton $\phi$ interacting with other fields
the evolution is governed by 
\begin{equation}
 \ddot{\phi} -\nabla^2\phi+ 3 H \dot{\phi} + \Upsilon \dot{\phi} + V' = (2\Upsilon T)^\frac12 \xi(x,t) \; ,
\label{phieom1}
\end{equation}
where, $H$ is the Hubble parameter during inflation, the overdots are derivatives w.r.t. cosmic time, prime indicates the derivative w.r.t. $\phi$, $\Upsilon$ is the inflaton dissipation rate and $\xi$ represents 
stochastic fluctuations due to interactions of $\phi$ with other species.  The stochastic r.h.s and
the dissipation rate on the l.h.s are related through the fluctuation-dissipation theorem
\cite{Graham:2009bf,Berera:2008ar,Ramos:2013nsa}. 
For the background homogeneous field $\phi(t)$, and a zero mean for $\xi(t)$
\begin{equation}
 \ddot{\phi} + 3 H \dot{\phi} + \Upsilon \dot{\phi} + V' = 0 \; .
\label{phieom}
\end{equation}

Recall that for a damped harmonic oscillator, $\ddot{x} + 2b \dot{x} + \omega^2 x = 0$, and the friction term, i.e. $2b\dot{x}$, causes dissipation and non-conservation of energy. Moreover, the rate of loss of energy is given by $dE/dt= -b \dot{x}^2$. Similarly, during warm inflation, for the scalar field we have
\cite{Mar-Berera-IJMP} 
\begin{equation}
\dot\rho_\phi+3H(\rho_\phi+p_\phi)=-\Upsilon{\dot\phi}^2,
\end{equation}
where $\Upsilon{\dot\phi}^2$ is the rate at which the inflaton loses energy.
Dissipation causes radiation to be produced and the radiation energy density $\rho_r$ evolves as
\begin{equation}
\dot\rho_r+4H\rho_r=\Upsilon{\dot\phi}^2~.
\label{rhordoteqn}
\end{equation}
Assuming the radiation thermalises quickly, the radiation energy density and temperature are related by
\begin{equation}
\rho_r=(\pi^2/30) g_* T^4  , 
\end{equation}
where, $T$ is the temperature of the radiation bath and $g_*$ is the number of relativistic degrees of freedom during warm inflation. The notion of thermal equilibrium and a temperature is only valid for $T_k>H_k$, as the Hubble radius $H_k^{-1}$ is then greater than the thermal de Broglie wavelength $\sim T_k^{-1}$  \cite{Imp}.   This then
gives a lower bound on $Q_P$, as we discuss later.

The various different realizations of warm inflation lead to different dependences of $\Upsilon$ on the temperature.
Here we focus on the models of warm inflation in which the inflaton is coupled only indirectly to 
light fields $(m<T)$ through heavy mediator fields $(m>T)$.
(Direct coupling of light fields to the inflaton gives large thermal
corrections to the potential thereby spoiling its flatness and effectively leading to too few e-foldings
of inflation \cite{BereraGleiserRamos1998,yokoyamalinde1999}.)
Then one finds that \cite{Moss:2006gt,BasteroGil:2010pb,BasteroGil:2012cm}
\begin{equation}
\Upsilon(\phi,T)=C_\phi \frac {T^3}{\phi^2} \; ,
\label{Upsilon}
\end{equation}
where $C_{\phi}$ is a dimensionless constant. 
Typically, in supersymmetric models such as the ones considered in 
Ref. \cite{BasteroGil:2012cm,Imp}
one gets
\begin{equation}
C_\phi = \frac{h^2}{16 \pi} N_X N_Y \; ,
\label{cphi_h}
\end{equation}
where $N_X$ and $N_Y$ are the number of heavy mediator fields and light fields respectively and $h$ is a Yukawa coupling.  
Note that for a small effective Yukawa coupling of $h\sqrt N_Y$ there can 
be a large  contribution to $\Upsilon$ from dissipation to on shell
heavy mediator particles, that later decay to Y particles, due to a narrow resonance that overcomes the Boltzmann suppression \cite{BasteroGil:2012cm}.

In the slow roll approximation, one can ignore the $\ddot{\phi}$ term, and
the background field evolution equation gives 
\begin{equation}
 \dot\phi\approx \frac{-V'(\phi)}{3H(1+Q)} \, ,
 \label{phidot}
\end{equation}
where $Q$ is defined by
\begin{equation}
Q=\frac{\Upsilon(\phi,T)}{3H(t)}.
\label{Qdefn}
\end{equation}
Similarly, ignoring $\dot{\phi}^2$ as compared to $V(\phi)$, and the sub-dominant $\rho_r$, 
Einstein's equation implies that 
\begin{equation}
H^2= \frac{8\pi}{3} \frac{V(\phi)}{\mpl^2} \; ,
\label{Hubble0}
\end{equation}
where $\mpl= 1/{\sqrt G} =1.2\times10^{19}$ GeV is the Planck mass. 
The slow roll parameters for warm inflation are defined as
\begin{equation}\label{slow_roll_parameters}
\epsilon_\phi = \frac{M_{Pl}^2}{16\pi}\,\left(\frac{V_{,\phi}}{V}\right)^2, \quad \eta_\phi = 
\frac{M_{Pl}^2}{8\pi}
\,\frac{V_{,\phi\phi}}{V},\quad\beta_\Upsilon = 
\frac{M_{Pl}^2}{8\pi}
\,\left(\frac{\Gamma_{,\phi}\,V_{,\phi}}{\Gamma\,V}\right)\,.
\end{equation}
In the literature, one also defines the Hubble flow function
\cite{Schwarz:2001vv}
\begin{equation}
\epsilon_1=\frac{d\ln(H_P/H)}{dN}=-\frac{\dot{H}}{H^2} \; .
\end{equation}
The slow roll parameters and the Hubble flow function are evaluated at the pivot scale.
Setting $\dot H=H_{,\phi}\dot\phi$ it is easy to verify that
\begin{equation}
\epsilon_1 = \frac{\epsilon_{\phi}}{1+Q}= \frac{1}{1+Q} \frac{\mpl^2}{16\pi} \left(\frac{V'}{V}\right)^2.
\end{equation}
The slow roll conditions needed for the inflationary phase are
\cite{Moss:2008yb}
\begin{equation} \label{slow_roll}
\epsilon_\phi \ll 1+Q,\quad |\eta_\phi| \ll 1+Q,\quad |\beta_\Upsilon| \ll 1+Q\,.
\end{equation}

In warm inflation one presumes that $\dot\rho_r\approx 0$ and the temperature of the Universe is approximately unchanged.
Then 
\begin{equation}
\rho_r\approx \frac{3}{4} Q {\dot\phi}^2\,.
\label{rhor}
\end{equation}
Figures (\ref{phiQkrho}) and (\ref{Tvsx}) show how $\rho_r$ and $T$ vary with time during warm inflation. $\dot\rho_r\approx 0$ is valid when modes of cosmological interest cross the horizon during inflation, but it seems to be only approximately true for the entire duration of inflation.
However $\dot\rho_r$ is small compared to the other terms in Eq. \eqref{rhordoteqn} throughout
inflation 
and hence the approximation is valid.

\section{Warm inflation model with a quartic potential}
\label{Modelandspectrum}

We consider a warm inflation model with a quartic potential 
\begin{equation}
V(\phi)=\lambda  \phi^4 \, 
\label{Vphi}
\end{equation}
and a dissipation coefficient $\Upsilon=C_\phi \frac{T^3}{\phi^2}$. Then
\be
Q=\frac{\Upsilon}{3H}=\frac{C_\phi}{3H} \frac{T^3}{\phi^2} \; .
\label{Q}
\ee
For a quartic potential, we get 
\be
H^2 = \frac{8\pi}{3} \frac{\lambda \phi^4}{\mpl^2}
\label{Hubble}
\ee
and the Hubble flow function
\begin{equation}
\epsilon_1 = \frac{1}{1+Q} \frac{1}{16\pi}  \left(\frac{\mpl}{\phi}\right)^2.
\end{equation}
It is easy to verify that $\epsilon_\phi = \frac{2}{3}\eta_\phi = -\beta_\Upsilon$.

\subsection{The primordial power spectrum}
\label{powerspectrum}

The scalar primordial power spectrum is given in 
Ref. \cite{Imp} (based on Refs. \cite{Berera:1995wh, Moss:1985wn, Berera:1999ws, Hall:2003zp, Ramos:2013nsa})
as 
\begin{equation}
P_\mathcal{R}(k)=\left(\frac{H_k^2}{2\pi\dot\phi_k}\right)^2\left[1+2n_k+\left(\frac{T_k}{H_k}\right)
\frac{2\sqrt{3}\pi Q_k}{\sqrt{3+4\pi Q_k}}\right]~,
\label{Pkfull}
\end{equation}
where $n(k)$ is the distribution of quanta of inflaton fluctuations $\delta \phi$ and the subscript $k$ refers to the time when a $k$ mode leaves the Hubble radius during inflation.
 Treating the inflaton field and the radiation
as coupled fluids, 
it has  been argued that 
perturbations in the radiation, because of inhomogeneous
dissipation, can lead to a growing mode in the inflaton perturbations which causes an enhancement in the primordial power spectrum in terms of a multiplicative factor whose form depends on the
	form of $\Upsilon$ \cite{BasteroGil:2011xd,Bastero-Gil:2016qru}. For a cubic dissipation coefficient, it is obtained numerically in Ref. \cite{Benetti:2016jhf}  as $G_{{\rm cubic}}(Q_P) = 1+ 4.981\  Q_P^{1.946}+ 0.127\  Q_P^{4.330}$. As our analysis is done for the weak dissipation regime, $10^{-5}<Q_P< 10^{-1}$, we can neglect it, as it is approximately 1 for the whole range of $Q_P$ of interest to us.
	(For large dissipation, $Q_P\gg1$, it was argued that  inflaton perturbations can be sufficiently 
enhanced to make
them incompatible with observations \cite{Graham:2009bf}, but subsequently it was shown that
shear viscosity in the non-equilibrium radiation damps the radiation 
perturbations and hence the enhancement of the inflaton perturbations.)  
The tensor power spectrum is given by the same expression as in cold inflation, namely,
\begin{equation}
 P_T(k) = \frac{16}{\pi} \left(\frac{H_k}{\mpl}\right)^2 \; .
 \label{PkT}
\end{equation}

We wish to express the scalar and tensor power spectra as functions of $k$. 
To this end, we look at the each term appearing in Eq. (\ref{Pkfull}) individually:
\begin{itemize}
 \item {\it Prefactor:} Using Eqs. (\ref{phidot}) and (\ref{Hubble})
\begin{equation}
\frac{H_k^2}{2\pi\dot\phi_k}=
\sqrt{\frac{8\pi}{3}}\sqrt{\lambda}
\left(\frac{\phi_k}{\mpl}\right)^3 (1+Q_k)\,.
\label{front}
\end{equation}

 \item {\it $T/H$ term:} Using the expression for $\rho_r(T)$ in Eq. (\ref{rhor}), and Eqs. (\ref{phidot}), (\ref{Vphi}) and (\ref{Hubble}) we get
\begin{equation}
T_k=\left(\frac{15}{\pi^3 g_*} \frac{Q_k}{(1+Q_k)^2} \lambda \,\phi_k^2 \mpl^2\right)^\frac{1}{4} \; .
\label{Tk}
\end{equation} 
Furthermore from Eq. (\ref{Hubble}) we obtain
\footnote{We note that Eq. (27) of Ref. \cite{Bart} gives another expression for ${T_k}/{H_k}$, which we do not use
	(it can be obtained by multiplying Eqs. (\ref{Hubble}) and (\ref{Qdefn}), and then using Eq. (\ref{Upsilon}) on the r.h.s.).}
\begin{equation}
\frac{T_k}{H_k}=\left(\frac{15}{\pi^3 g_*}\right)^\frac{1}{4}\sqrt{\frac{3}{8\pi}}\lambda^{-\frac{1}{4}}
 \frac{Q_k^{\frac{1}{4}}}{(1+Q_k)^{\frac{1}{2}}} 
\left( \frac{\phi_k}{\mpl}\right)^{-\frac{3}{2}} \; .
\label{TkoverHk}
\end{equation}

 \item {\it Occupation number:} We take the inflaton distribution, $n_k$, to be represented by a Bose-Einstein distribution,
\begin{equation}
n_k=\frac{1}{\exp(\frac{k/a_k}{T_k}) -1}
\label{BE}
\end{equation}
where $a$ is the scale factor and $T$ represents the physical temperature.  Then the first two terms in the square bracket of Eq. (\ref{Pkfull}) become
\begin{equation}
1+2n_k= \coth\frac{H_k}{2T_k}\,,
\label{coth}
\end{equation}
with $H_k/T_k$ obtained from Eq. (\ref{TkoverHk}).
If 
$n_k=0$ then 
\begin{equation}
1+2n_k\rightarrow 1\,.
\label{nocoth}
\end{equation}
We discuss the importance of the $\coth$ term in Sec. \textsection \ref{cothsection}.

\end{itemize}

The above discussion implies that 
the scalar power spectrum in Eq. (\ref{Pkfull}) 
depends on 
$\lambda$, $Q_k$ and $\phi_k$. 
Furthermore, using the slow roll equations, Eqs. (\ref{phidot}) and (\ref{rhor}),
and  Eq. \eqref{Upsilon} for $T$,
one can express
$Q_k$ as a function of $\phi_k$.  
This can then be inverted
 to give
 $\phi_k$ in terms of $Q_k$, as in Eq. (21) of Ref. \cite{Bart}, as \footnote{ Note that
	our $\lambda$ is equivalent to $\lambda^2$ in Ref. \cite{Bart}.}
\begin{equation}
 \left(\frac{\phi_k}{\mpl}\right)=\sqrt\frac{1}{8\pi}\left(\frac{64 C_{\phi}^4\lambda}{9 C_R^3}\frac{1}{Q_k(1+Q_k)^6}\right)^{\frac{1}{10}}~,
 \label{phikovermpl}
\end{equation}
where $C_R=(\pi^2/30)g_*$. 

\subsection{Obtaining the dissipation parameter $Q_k$}

In order to obtain the $k$ dependence of the scalar power spectrum, 
we need to find the $k$ dependence of $Q_k$. From  
Eq. (13) of Ref. \cite{Imp} (based on Ref. \cite{Bart}) we have
\footnote{Our expression for $dQ/dN$ differs from that in 
Ref. \cite{Imp} by a minus sign because our $N$ is measured from the end of inflation.} 
\begin{equation} 
\frac{dQ}{dN}= -\frac{1}{C_1}\frac{Q^{6/5}(1+Q)^{6/5}}{1+7Q}~,
\label{dQdN}
\end{equation}
where $N = \ln (a_e/a)$ is the number of e-foldings of inflation from the end of inflation, and 
\begin{equation}
 C_1 = \frac{1}{40}\left(\frac{64C_{\phi}^4\lambda}{9C_R^3}\right)^{\frac15}~.
 \label{C1}
\end{equation}
In order to find the $k$ dependence, it is useful to define $x=\ln(k/k_P)$. Then, 
\begin{equation}
\frac{ dQ}{dx} =\frac{ dQ}{dN}\frac{ dN}{dx} \; .
\label{dQbdx}
\end{equation}
Since $N= \ln (a_e/a)$ and $a=k/H$, $dN/dx=-1+H^{-1}dH/dx=
-1+H^{-1}(dH/dN)\, (dN/dx)$. Then using $\epsilon_1=H^{-1}dH/dN$, we get
\begin{equation}
\frac{dN}{dx}= -\frac{1}{1-\epsilon_1} 
\label{dNdx}\,.
\end{equation}

Therefore 
\begin{equation}
\frac{ dQ}{dx}=\frac{1}{[1-\epsilon_1]}\frac{1}{C_1}\frac{Q^{6/5}(1+Q)^{6/5}}{1+7Q} \; .
\label{dQkbydx}
\end{equation}
One integrates the above equation from the time when $k^{th}$ mode leaves the horizon till when the pivot scale crosses the horizon 
to get a relation between
$Q_k$ and $k$ in the form of a transcendental equation as 
\be
G(Q_k) - G(Q_p) = x\,,
\label{GQx}
\ee
where
\begin{align}
G(Q) &= \frac{C_1}{12Q^{1/5}(1+Q)^{6/5}}\left[-32\left(\frac{10125 \pi^6 Q(1+Q)^6}{C^4_\phi \lambda}\right)^{1/5}\ln \left( Q(1+Q)^6\right) \right. \nonumber\\  
&+ \left.   15(1+Q)\left(-4+ 20Q -15Q(1+Q)\, _2F_1\left(1,\frac{8}{5},\frac{9}{5},-Q\right) \right) \right]
\label{GQ}
\end{align}
and $_2F_1 (a, b, c, z)$ is the Hypergeometric function.
 
Combining Eqs.  \eqref{GQx} and \eqref{GQ} 
with
the expressions in Sec. \textsection \ref{powerspectrum}
we can express the power spectrum $P_\mathcal{R}$ as a function of
$k$, and the parameters $\lambda, C_\phi$ and $Q_P$.

\subsection{Duration of inflation}
\label{durationofinflation}

We have to ensure that the parameters we choose for the power spectrum correspond to a scenario that
gives sufficient e-foldings of inflation.  
Applying the condition that inflation lasts for $N_P$ e-foldings after the pivot scale crosses the horizon during inflation provides
a relation between the parameters $\lambda$ and $C_\phi$. 
Integrating Eq. \eqref{dQdN} from the time when $k^{th}$ mode leaves the horizon till when the pivot scale crosses the horizon  we get
\footnote{
Note that $C_1$ is referred to as $C_Q$ in  Ref. \cite{Bart} and 
$1/C_Q$ in Ref. \cite{Mar-Berera-IJMP} while
$1/C_1$ is $C_*$ in Ref. \cite{Imp} which is approximated as
$
C_1^{-1}=C_* \approx 5\epsilon_k Q_k^{-1/5} \,\,\, {\rm for}\,\,\,  Q_k\ll1
\label{C1_1}
$, 
where $\epsilon_k$ is a $k$ dependent slow roll parameter.   
(On substituting the definition of $\epsilon_\phi$,
$C_*= 5\epsilon_\phi Q_P^{-1/5}= 5{M_{Pl}^2}/{16\pi}\left({V'}/{V}\right)^2_P Q_P^{-1/5} 
= 40\left(\frac{M_{Pl}}{\sqrt{8\pi}\phi_P}\right)^2 Q_P^{-1/5}$.
Using Eq. (\ref{phikovermpl}), 
this is equal to
$40 \left(\frac{9C_R^3 Q_P (1+Q_P)^6}{2^6 C_\phi^4 \lambda} \right)^{2/10} Q_P^{-1/5}= 40\left(\frac{9C_R^3}{2^6 C_\phi^4 \lambda} \right)^{1/5}(1+Q_P)^{6/5} \approx 40\left(\frac{9C_R^3}{2^6 C_\phi^4 \lambda} \right)^{1/5}= \frac{1}{C_1}$ 
for $Q_P\ll1$.
Hence $C_*$ is independent of $k$ for $Q_P\ll 1$.)
}   
\be
N_k-N_P=C_1[F(Q_P)-F(Q_k)]\,,
\label{NP}
\ee
where $C_1$ is as given in Eq. \eqref{C1} and 
\be
F(Q)=\frac{5}{4 Q^{\frac{1}{5}}(1+Q)}
\left[
4(1+Q)^\frac{4}{5}(-1+5Q)-15Q(1+Q) \cdot
~_2F_1\left( \frac{1}{5},\frac{4}{5},\frac{9}{5},-Q \right)\,
\right]\,,
\label{FQ}
\ee
where again $_2F_1 (a, b, c, z)$ is the Hypergeometric function.
\footnote{Note that our $F(Q)$ is different from $F_r(Q)$ in Eq. (26) of Ref. \cite{Bart}. 
We are unable to reproduce Eq. (25) of Ref. \cite{Bart}.
}

Let us consider the mode which crosses the Hubble radius at the end of inflation.
For such a mode $N_k = 0$ and Eq. (\ref{NP}) implies that
\be
N_P=C_1[F(Q_e)-F(Q_P)]\,.
\label{NP2}
\ee
We can obtain $Q_e$ by setting the second potential slow roll parameter 
$\eta_\phi$ to $1+Q_e$, i.e.
\be
\eta_e=\frac{3}{2\pi}\frac{\mpl^2}{\phi_e^2}=1+Q_e\,.
\ee
For the model under consideration, warm inflation ends because the slow
roll conditions fail, and not because $\rho_r$ overtakes $\rho_\phi$, as seen in Figure (\ref{phiQkrho}), 
where $\rho_{\phi e}\sim3\rho_{r e}$.
(Since, as shown earlier, $\eta_\phi$ is the largest slow roll parameter it is sufficient to verify
the breakdown of the slow roll condition with $\eta_\phi$.)
Expressing $\phi_e$ in terms of $Q_e$ 
using Eq. (\ref{phikovermpl}) we 
get 
\beq
Q_e^2 + Q_e - \left(\frac{64C_{\phi}^4\lambda}{9C_R^3}\right)\frac{1}{12^5}=0\,,
\eeq
whose positive solution is
\beq
Q_e=\frac{-1+\sqrt{1+4\left(\frac{64C_{\phi}^4\lambda}{9C_R^3}\right)\frac{1}{12^5}}}{2}\,.
\label{Qe}
\eeq
Notice that $Q_e$ depends on $\lambda C_\phi^4$.

Using the expressions for $C_1$ (Eq. \eqref{C1}) and $Q_e$ (Eq. \eqref{Qe}) in Eq. (\ref{NP2}), 
we get a value for $\lambda C_\phi^4$ for a given value of $N_P$ and $Q_P$ which we can use to express $C_\phi$ in terms of $\lambda$.
Using this in $P_\mathcal{R}(k)$ the parameters defining the power spectrum are now $\lambda, Q_P$ and $N_P$.  Similarly one can obtain
the tensor power spectrum $P_T(k)$ in terms of these parameters.
By choosing the value
of $N_P$ appropriately to be, say 50 or 60, we can ensure that we have a
sufficient duration of inflation.

Thus, the above implies that it is the values of $\lambda, Q_P$ and $N_P$ that need to be constrained from the CMB observational data.  Therefore
in our {\tt CosmoMC} code we vary the parameters $\lambda$ and $Q_P$ while
keeping $N_P$ fixed for a particular run.

As an aside we note that,
\bea
N_k&=&\ln \frac {a_e}{a_k}=\ln \frac {a_e}{a_P}+\ln \frac {a_P}{a_k}\cr
&=&N_P+\ln \left( \frac{k_P}{k}\frac{H_k}{H_P}\right)\cr
&=&N_P+\ln \frac{k_P}{k} + \ln\frac{\phi^2_k}{\phi^2_P}\,,
\eea
which, using Eq. (\ref{phikovermpl}), gives
\be
N_k-N_P=-\ln \frac{k}{k_P} + \ln\left[\frac{Q_P(1+Q_P)^6}{Q_k(1+Q_k)^6}\right]^\frac{1}{5} \; .
\label{NkNp}
\ee
Combining Eqs. (\ref{NP}) and (\ref{NkNp}) then gives us $Q_k$ as a function
of $\ln(k/k_P)$, for a given $Q_P$ and $N_P$.  We have numerically confirmed that this
agrees with the $Q_k(x)$ that we obtain from Eqs. \eqref{GQx} and \eqref{GQ}.  In our numerical
code we use Eqs. \eqref{GQx} and \eqref{GQ}. 

\section{Primary analysis using the primordial power spectrum and its normalisation}
\label{powerspectrumanalysis}

\begin{figure}[h]
\includegraphics[width=0.5\textwidth]{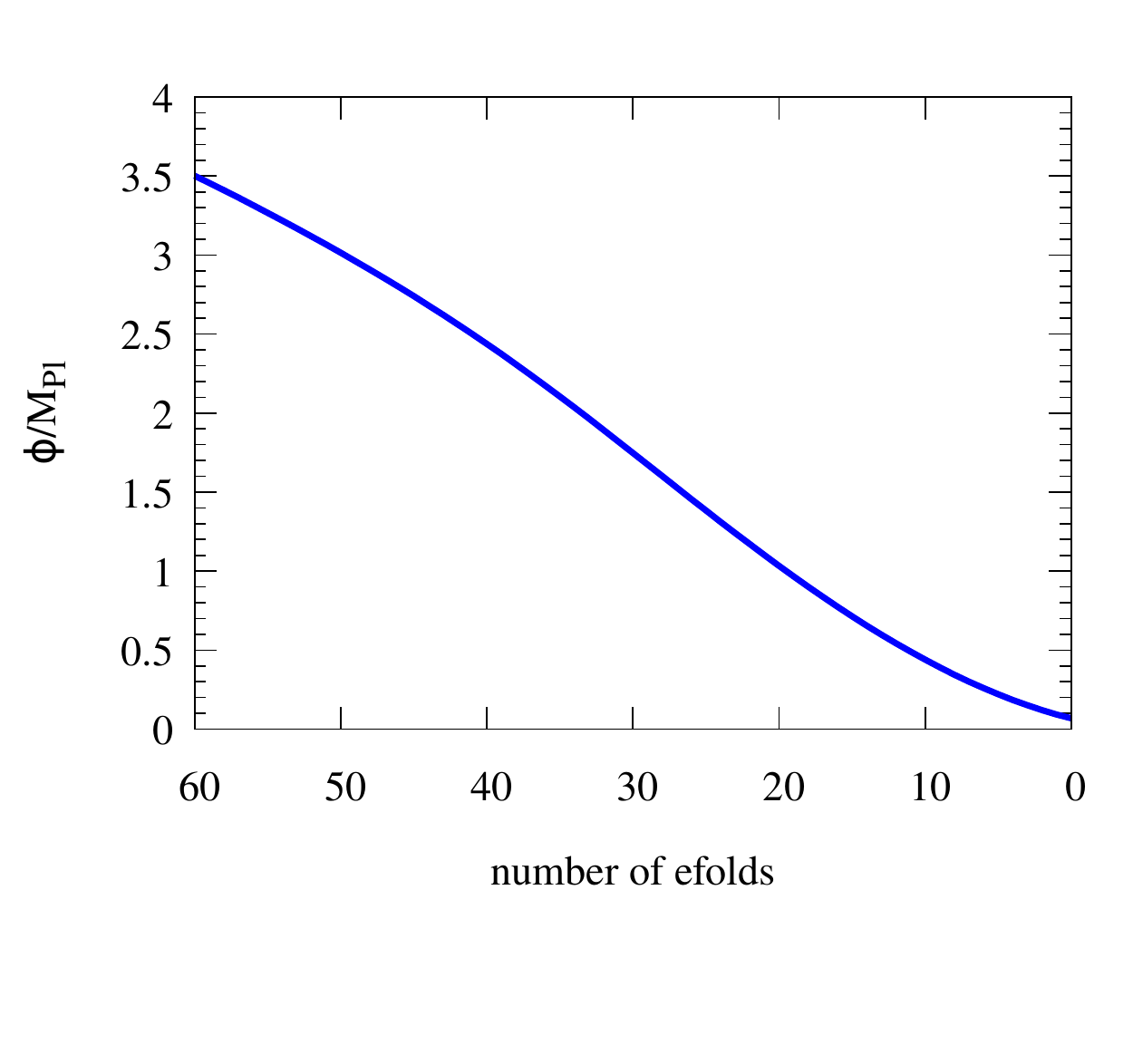}
 \includegraphics[width=0.5\textwidth]{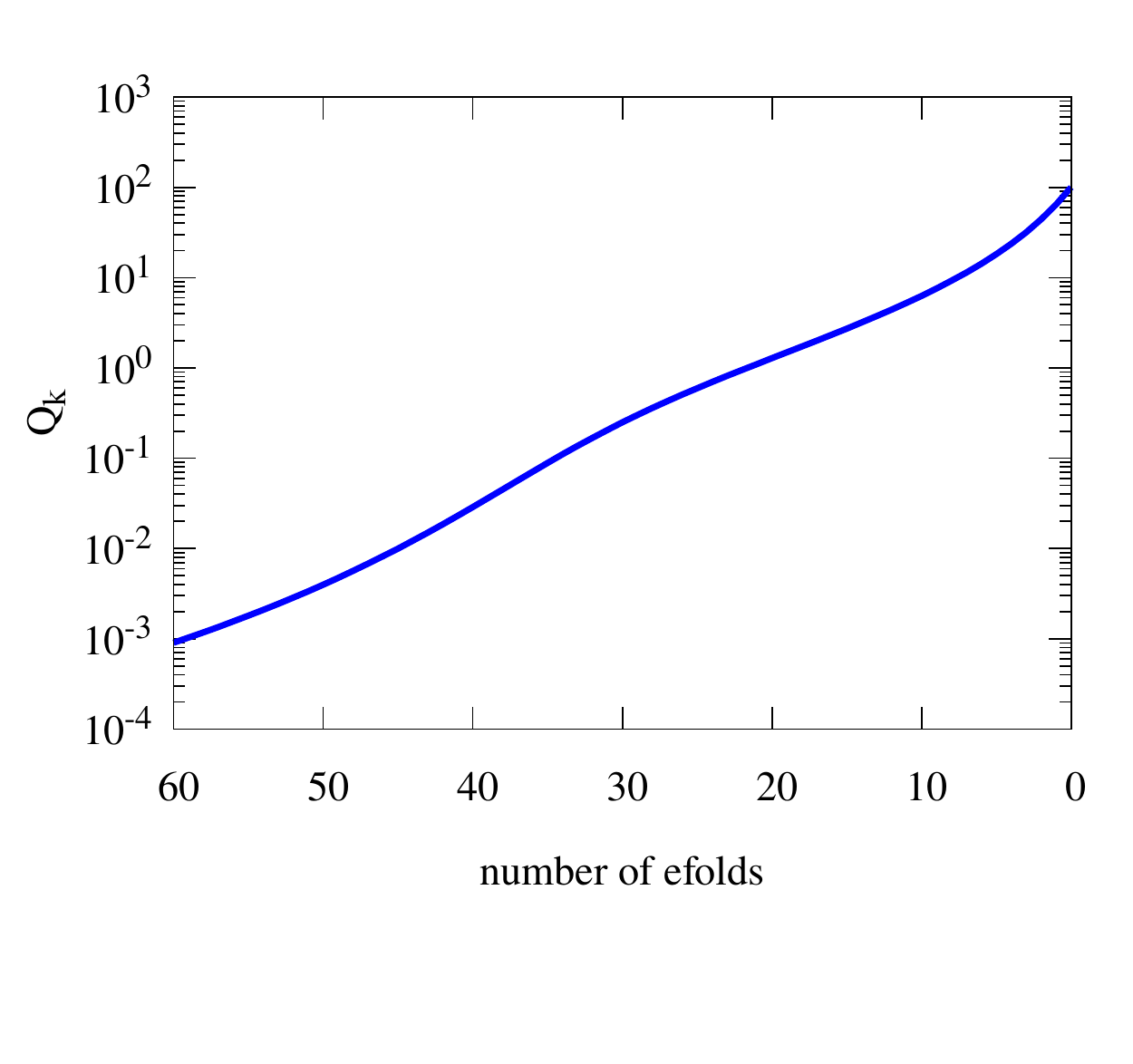}
 \vspace{-2cm}
\begin{center}
 \includegraphics[width=0.5\textwidth]{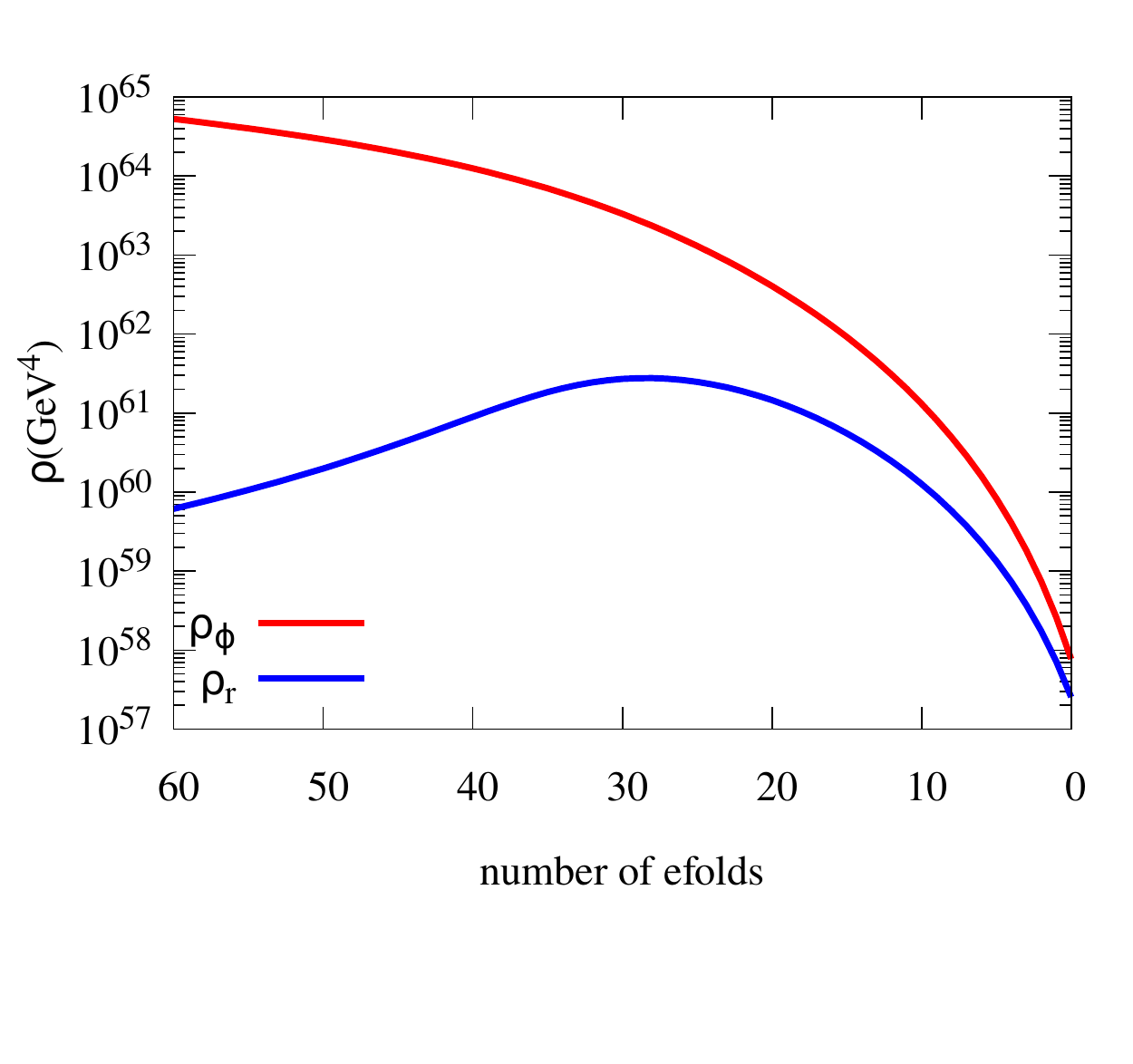}
 \end{center}
  \vspace{-1.8cm}
 \caption{
 The behaviour of the homogeneous inflaton field $\phi$, $Q_k$, and the energy density in
 $\phi$ and radiation is shown as a function of the number of efoldings $N$.  
 Note that as $Q$ crosses 1 during inflation the model
 transits from weak dissipation to strong dissipation.  For our model $\rho_\phi$ is larger than
 $\rho_r$ till when the slow roll conditions fail and the latter therefore
 determines when inflation ends.
 Consequently the Universe will also go through a phase of reheating after inflation.  All plots correspond 
 to $N_P=50$ and fixed $A_s$,  and $Q_P= 10^{-2.4}$.  
 }
  \label{phiQkrho}
\end{figure}

In the last section we argued that the primordial power spectrum 
is completely determined by the values of the three parameters $N_P$, $Q_P$ and $\lambda$.
These will be the parameters we vary when we perform a detailed data analysis with 
{\tt CosmoMC}.  But before presenting those results, 
we shall study using Mathematica 
the power spectrum and understand how these parameters determine 
various features of the quartic model under consideration.
For this section
we also impose the normalisation of the power spectrum 
as $P_\mathcal{R}(k_P)\equiv A_s=2.2\times
10^{-9}$.  For any given value of $Q_P$ and $N_P$ this constraint then fixes $\lambda$.
\footnote{For our subsequent {\tt CosmoMC} analysis we prefer $\lambda, Q_P$ and $N_P$ as the independent parameters as $\lambda$ is more fundamental to the model than $A_s$.}
To appreciate how parameters change during warm inflation we also plot 
the homogeneous inflaton field $\phi$, $Q_k$ and the energy density of $\phi$ and the radiation as a function
of the number of efoldings of inflation in Figure (\ref{phiQkrho}).
\vspace{-0.5cm}

\vspace{0.5cm}
\subsection{Spectral index}

\begin{figure}[h]
\centering
    \includegraphics[width=0.55\textwidth]{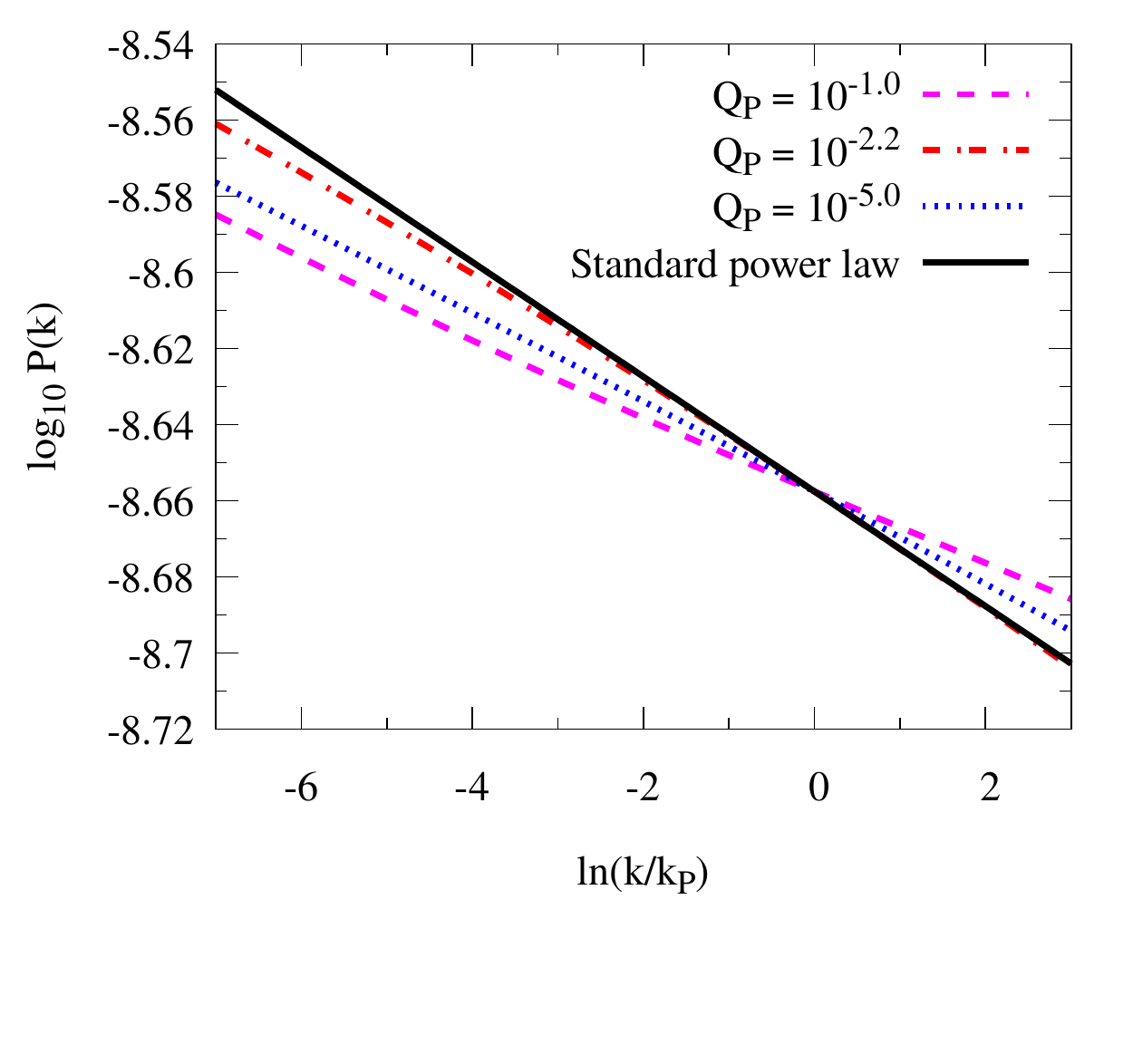}
    \vspace{-1.2cm}
  \caption{
  The primordial power spectrum, ${\rm{log}}_{10}P_\mathcal{R}(k)$ vs $\ln(k/kp)$ is shown for
 different values of $Q_P$ for $N_P = 50$ and $A_s$ fixed.  The coupling 
 $\lambda$
 is different for different values of $Q_P$.  The standard power law with $n_s=0.9653$ is also shown as a solid (black) line. 
} 
  \label{fig:powerspectrum50}
\end{figure}

\begin{figure}[h]
\includegraphics[width=0.5\textwidth]{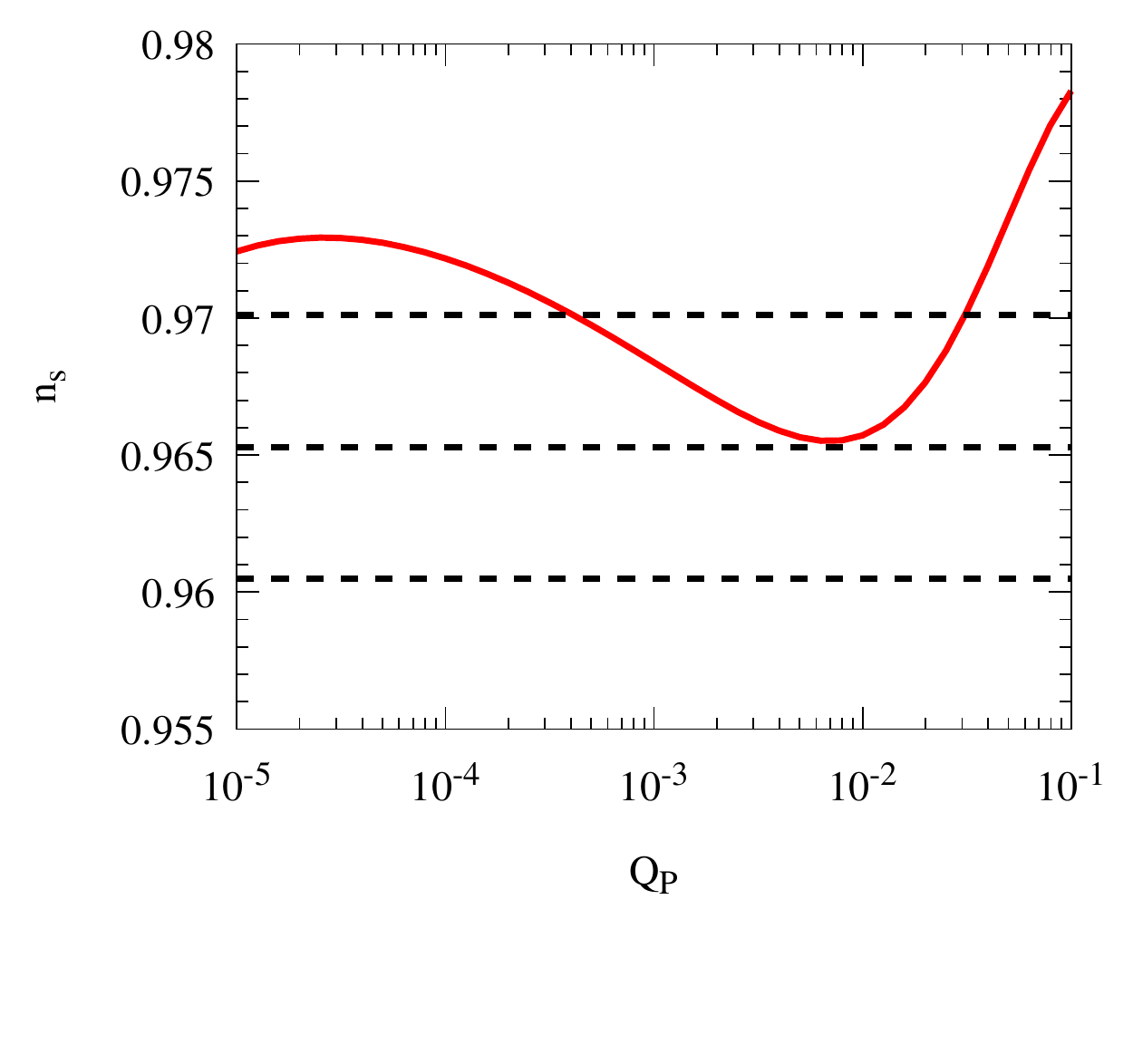}
 \includegraphics[width=0.5\textwidth]{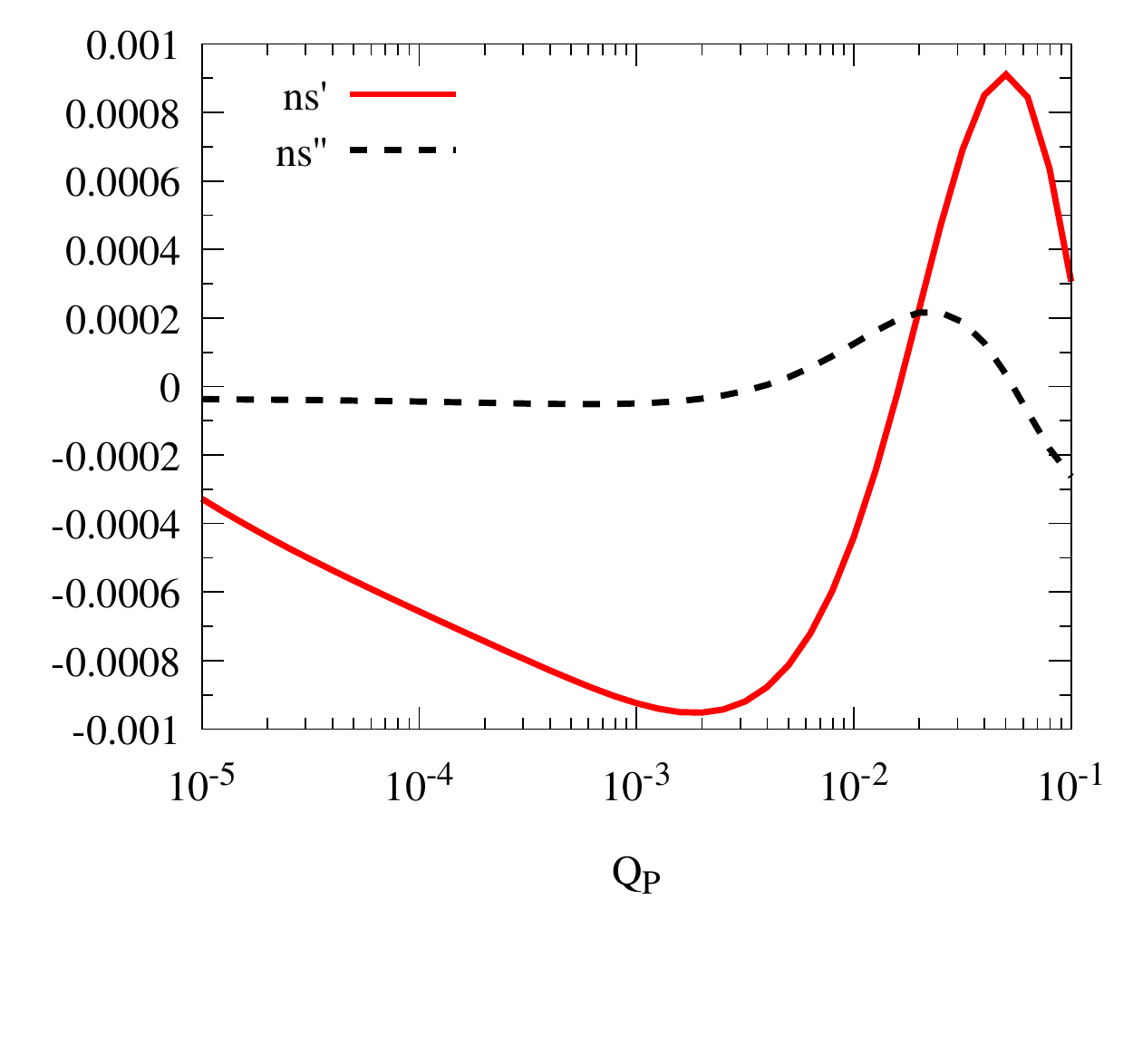}
 \vspace{-2cm}
 \caption{The behaviour of the scalar spectral index, $n_s$, as we change $Q_P$ for $N_P=50$ and $A_s$ fixed is shown by the solid curve in the first panel. The three horizontal dashed lines show the mean value and 68 $\%$ CL limits on $n_s$ for Planck 2015 TT, TE, EE + low P + lensing data for power law primordial power spectrum $(0.9653 \pm 0.0048)$ \cite{planck2015:params}.  Note the non-monotonic behaviour of $n_s$. 
 The running of the spectral index $n_s'\equiv dn_s/d(\ln k)$ and the running of the running $n_s'' \equiv d^2 n_s/d(\ln k)^2$ as a function of $Q_P$ are shown in the second panel for $N_P=50$. 
}
  \label{fig:dpkQp}
\end{figure}

We first plot the power spectrum 
for different values of $Q_P$ while keeping $N_P$ fixed at $50$ and $A_s$ fixed as above. 
This is shown in Figure (\ref{fig:powerspectrum50}),
along with the standard power law power spectrum with $n_s$ equal to the mean value
of $0.9653$
obtained by the Planck collaboration with TT, TE, EE + low P + lensing 2015 data \cite{planck2015:params}.
The different plots for warm inflation correspond to
different values of $\lambda$, chosen appropriately to give the correct normalisation $A_s$.
It is noteworthy that as we decrease $Q_P$ from $10^{-1}$ to $10^{-2.2}$, the slope of the power spectrum decreases,  and then increases as we further decrease $Q_P$ to $10^{-5}$.
This behaviour is best understood by studying the dependence of $n_s$ on $Q_P$ in Figure (\ref{fig:dpkQp}).
It is also worth noting from Figure (\ref{fig:dpkQp}) that for none of the values of $Q_P$ in the range we shall study will the power spectrum fall on top of the standard power law spectrum (the solid line in Figure (\ref{fig:powerspectrum50})). 

One can determine the scalar spectral index as
\begin{equation}
\left.n_s-1\right.
\equiv
 \left.\frac{d\ln P_\mathcal{R}(k)}{d\ln(k/k_P)}\right\rvert_{k=k_P}=\left.\frac{d\ln P_\mathcal{R}}{d Q_k}\frac{dQ_k}{d\ln(k/k_P)}\right\rvert_{k=k_P} \;.
\end{equation}
We obtain the first derivative by 
expressing $P_\mathcal{R}(k)$ in terms of $Q_k$ using Eqs. (\ref{Pkfull},\ref{front},\ref{TkoverHk},\ref{coth}) and (\ref{phikovermpl}) and then by taking the derivative w.r.t. $Q_k$.
$\frac{dQ_k}{d\ln(k/k_P)}$ is given in Eq. (\ref{dQkbydx}). Then the scalar spectral index can be written as 
\vspace{-0.5cm}
\begin{align}
n_s = &1 -\frac{3\epsilon_1}{1-\epsilon_1} 
+ \frac{10 \epsilon_1 \ Q_P}{(1+7 Q_P)(1-\epsilon_1)}
 \nonumber\\
&
+\frac{1}{(1-\epsilon_1)} \frac{4 n_P \epsilon_1 }{1+2 n_P+\delta} \left(\frac{e^{H_P/T_P}}{e^{H_P/T_P}-1} \frac{H_P}{T_P}\right) \left(\frac{1+2Q_P}{1+7Q_P} \right) \nonumber \\
&+\frac{1}{(1-\epsilon_1)} \frac{\delta \epsilon_1}{(1+ 7 Q_P)}\frac{1}{1+2 n_P+ \delta} 
 \left[ 2+4 Q_p + 5(1+Q_P) \left( \frac{3+2\pi Q_P}{3+4\pi Q_P} \right)\right]  \,,
\label{nsfull}
\end{align}
where $n_P=n(k_P)_{k_P}$ and we define $\delta=\frac{2 \sqrt{3} \pi Q_P}{\sqrt{3+4\pi Q_P}}\frac{T_P}{H_P}$.
\vspace{0.5cm}

The behaviour of $n_s$ as we change $Q_P$ is shown in Figure (\ref{fig:dpkQp}) which explains the changes in the slope of the power spectrum seen in Figure (\ref{fig:powerspectrum50}). 
Note that the value of $n_s$ that corresponds to the mean value of $Q_P$ of $10^{-2.4}$, 
obtained by us using {\tt CosmoMC}, is 0.9659 which differs slightly from the mean value of $n_s$ of  0.9653.
We also show the running of the spectral index at the pivot scale, and the running of the running,  as we vary $Q_P$ 
in the right panel of Figure (\ref{fig:dpkQp}) where we plot $dn_s/d(\ln k)$  and $d^2 n_s/d (\ln k)^2$ as a function of $Q_P$ for $N_P$ fixed at $50$.  The extremely weak running justifies our use in Figure (\ref{fig:dpkQp}) of the allowed band of $n_s$ values from the Planck collaboration for fixed $n_s$ without running.

\subsection{Coupling parameters $\lambda$ and $C_\phi$ as a function of $Q_P$}
\label{couplings}

\begin{figure}[h]
\includegraphics[width=0.5\textwidth]{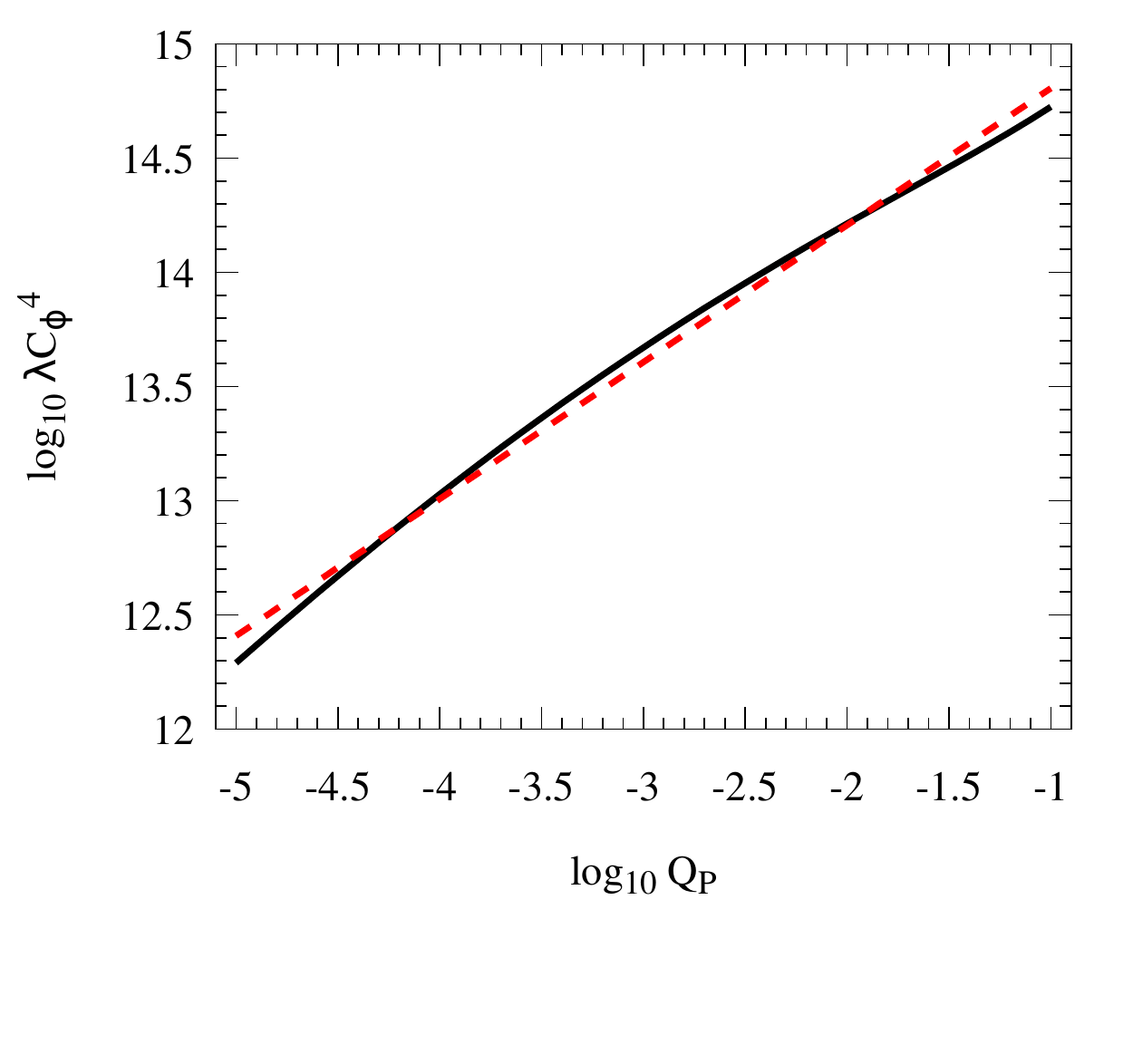}
\vspace{-1.4cm}
\includegraphics[width=0.5\textwidth]{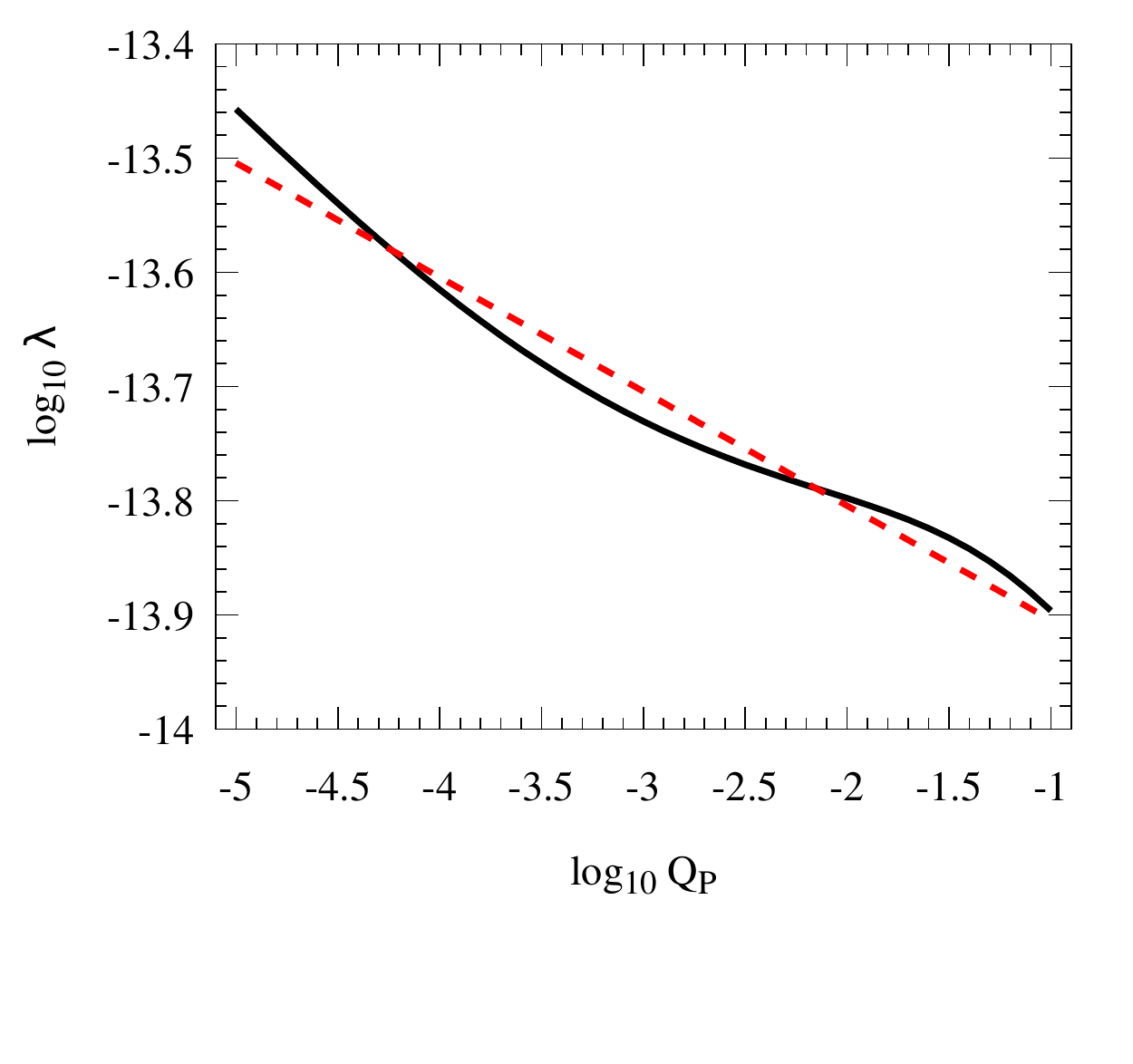}
 \includegraphics[width=0.5\textwidth]{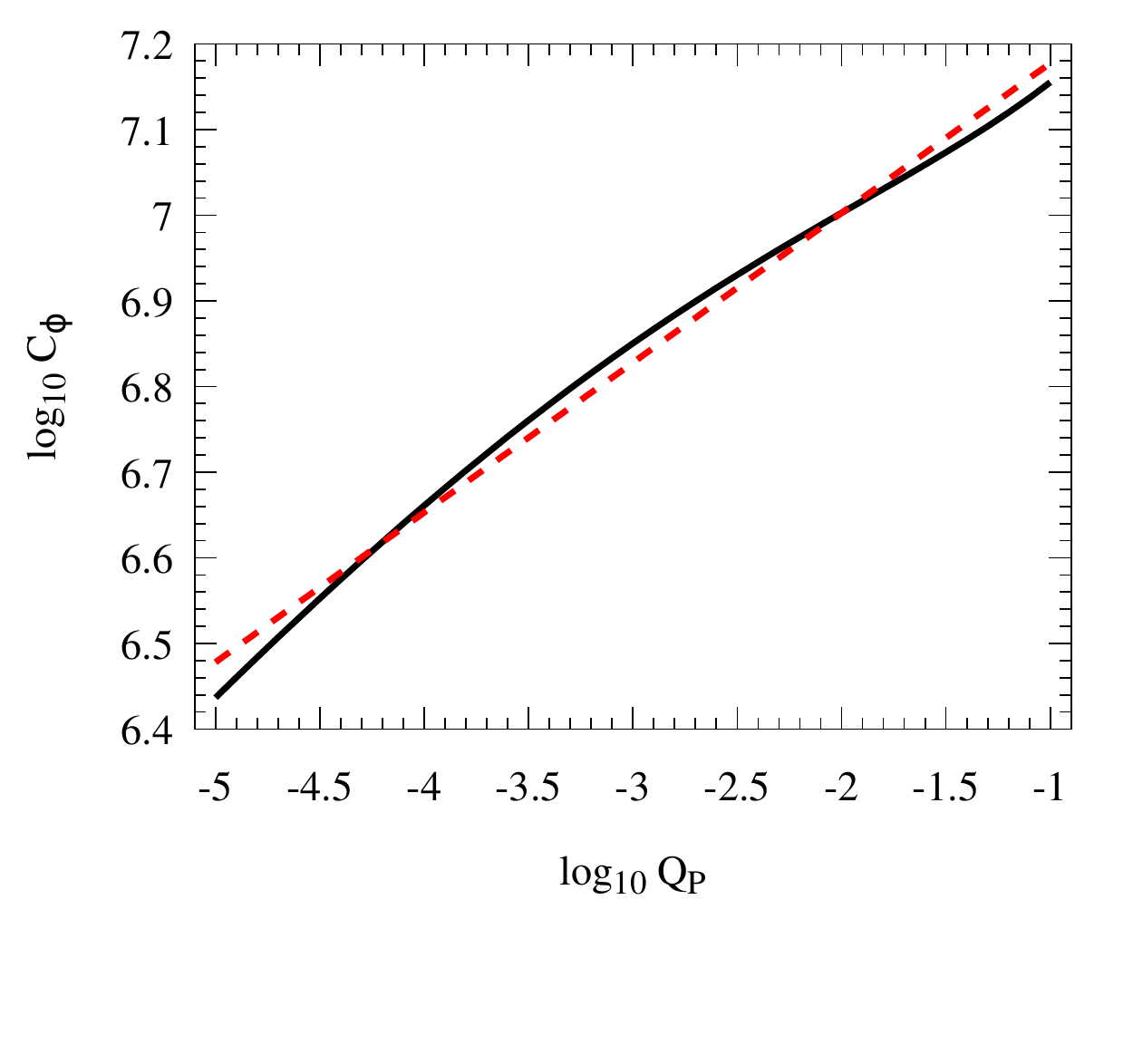}
   \includegraphics[width=0.5\textwidth]{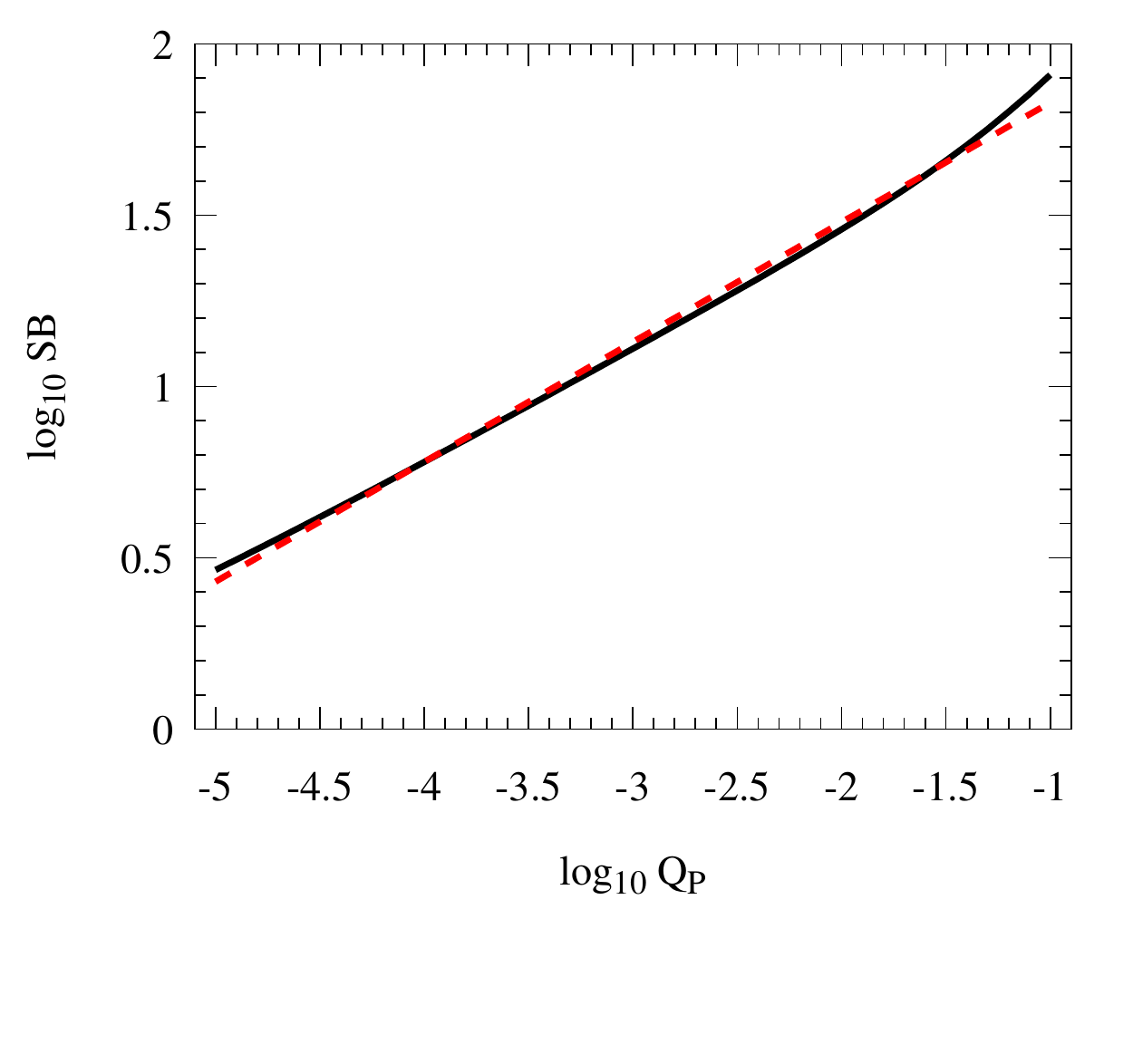}
   \vspace{-1.5cm}
  \caption{
  The first panel shows  $\lambda {\rm C}_{\phi}^4$ 
  as a function of $Q_P$ for $N_P=50$.  The next two panels show
  $\lambda$ and $C_\phi$ as functions of $Q_P$ for $N_P=50$ and
  $A_s=2.2\times10^{-9}$.  The last panel shows the square bracket in
  Eq. \eqref{Pkfull}  as a function of $Q_P$. In all the plots, the black solid curves correspond to the 
 numerical values of the corresponding quantity and the red dashed curves are straight line fits to the log-log plots. 
 For the first panel the fitting function is $\lambda {\rm C}_{\phi}^4 = Q_P^{0.60} \times 10^{15.45}$,  for the 
 next two plots they are $C_{\phi} = Q_P^{0.17}\times 10^{7.35}$
 and $\lambda = Q_P^{-0.10}\times 10^{-14.0}$, and for the final plot $SB = Q_P^{0.35}\times 10^{2.18}$.
  }
\label{fig:cphivsQP}
\end{figure}

As discussed in Sec. \textsection \ref{durationofinflation}, $N_P$ and $Q_P$ determine
the value of $\lambda C_\phi^4$.  
For $N_P=50$, we vary $Q_P$ and obtain log$_{10} \lambda C_\phi^4$ against $Q_P$, as shown in 
Figure (\ref{fig:cphivsQP}).  Fitting the log-log plots to a straight line, we get $\lambda {\rm C}_{\phi}^4 = Q_P^{0.60}\times 10^{15.45}$.
Imposing the normalisation $A_s$
of the power spectrum fixes the value of $\lambda$, as mentioned above,
and we obtain $\lambda$ and $C_\phi$ as functions of $Q_P$
in the other panels in Figure (\ref{fig:cphivsQP}).
The corresponding fitting functions are
 $C_{\phi} = Q_P^{0.17}\times 10^{7.35}$
 and $\lambda = Q_P^{-0.10}\times 10^{-14.0}$.
This then implies $\lambda\sim C_\phi^{-0.57}$.
\vspace{-0.2cm}

The correlation between $C_\phi$ and $Q_P$
tells us how the observational limits on $Q_P$, the decay width of the inflaton in units of thrice the Hubble parameter can help us estimate $C_\phi$. As one can see from Eq. (\ref{cphi_h}), $C_\phi$ is a function
of the Yukawa coupling $h$ and the number of mediators and light fields. As mentioned earlier, 
in the weak dissipation regime, the mean value of $Q_P$ that we obtain using CMB observations is of the order of $10^{-2.4}$. The plot of $C_\phi$ vs $Q_P$ in 
Figure (\ref{fig:cphivsQP}) then indicates that the corresponding value of $C_\phi$ is of the order of $8.8 \times10^{6}$. For values of the Yukawa coupling in the regime of validity of perturbative calculations, this implies that 
$N_X N_Y$ is of order a 100 million.
Thus, this implementation of warm inflation in which the inflaton couples to light degrees of freedom forming radiation through heavy mediator fields requires the existence of 
order 100 million fields in the corresponding microscopic theory.
This requirement of a large value of $C_\phi$,
or a large number of fields, has been 
pointed out in Ref. \cite{Berera:1998px, Mar-Berera-IJMP,Imp}.

\subsection{Thermal equilibrium and a lower bound on $Q_P$}
\label{sec:TgtH}

\begin{figure}[h!]
\includegraphics[width=0.5\textwidth]{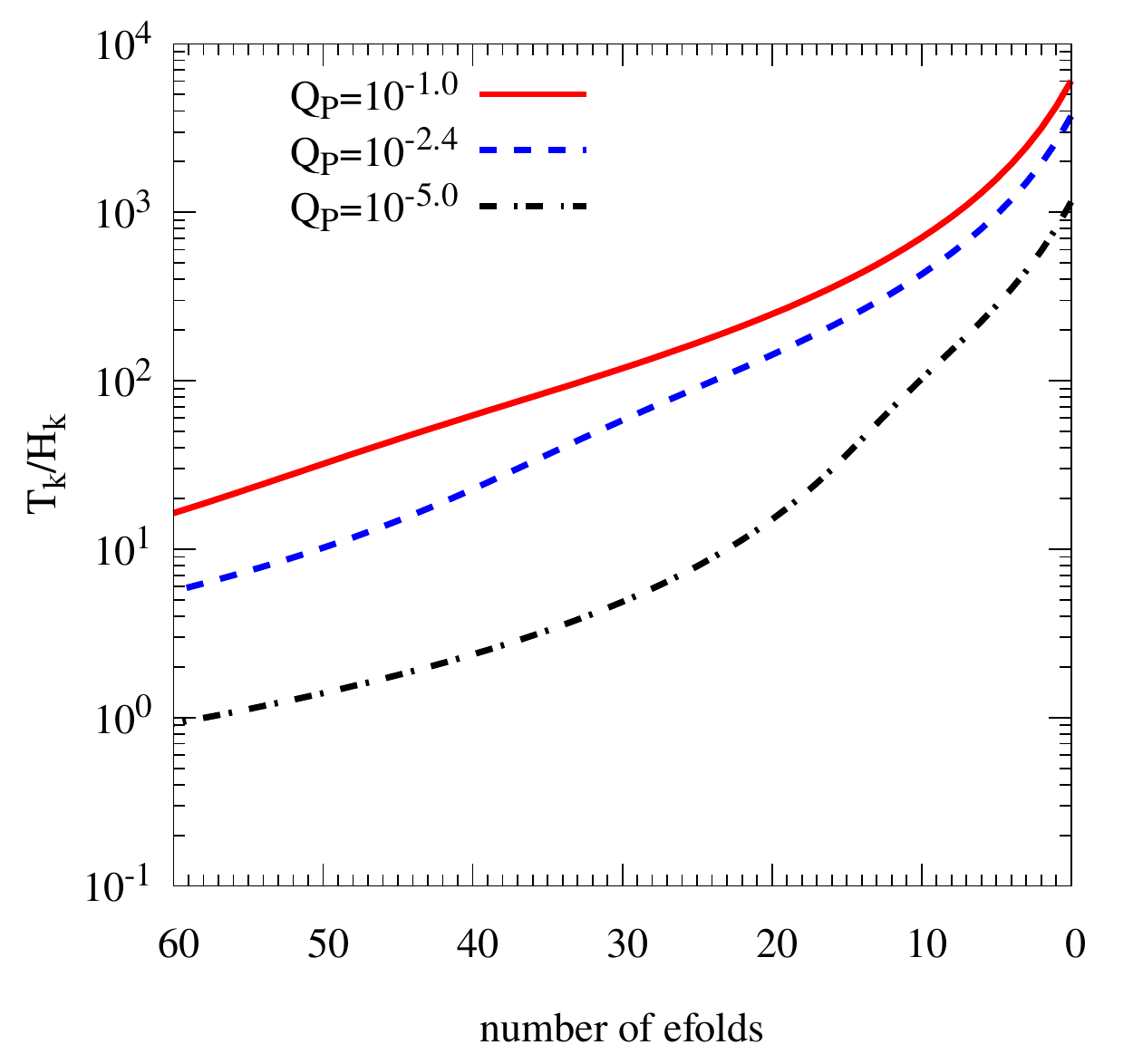}
  \includegraphics[width=0.5\textwidth]{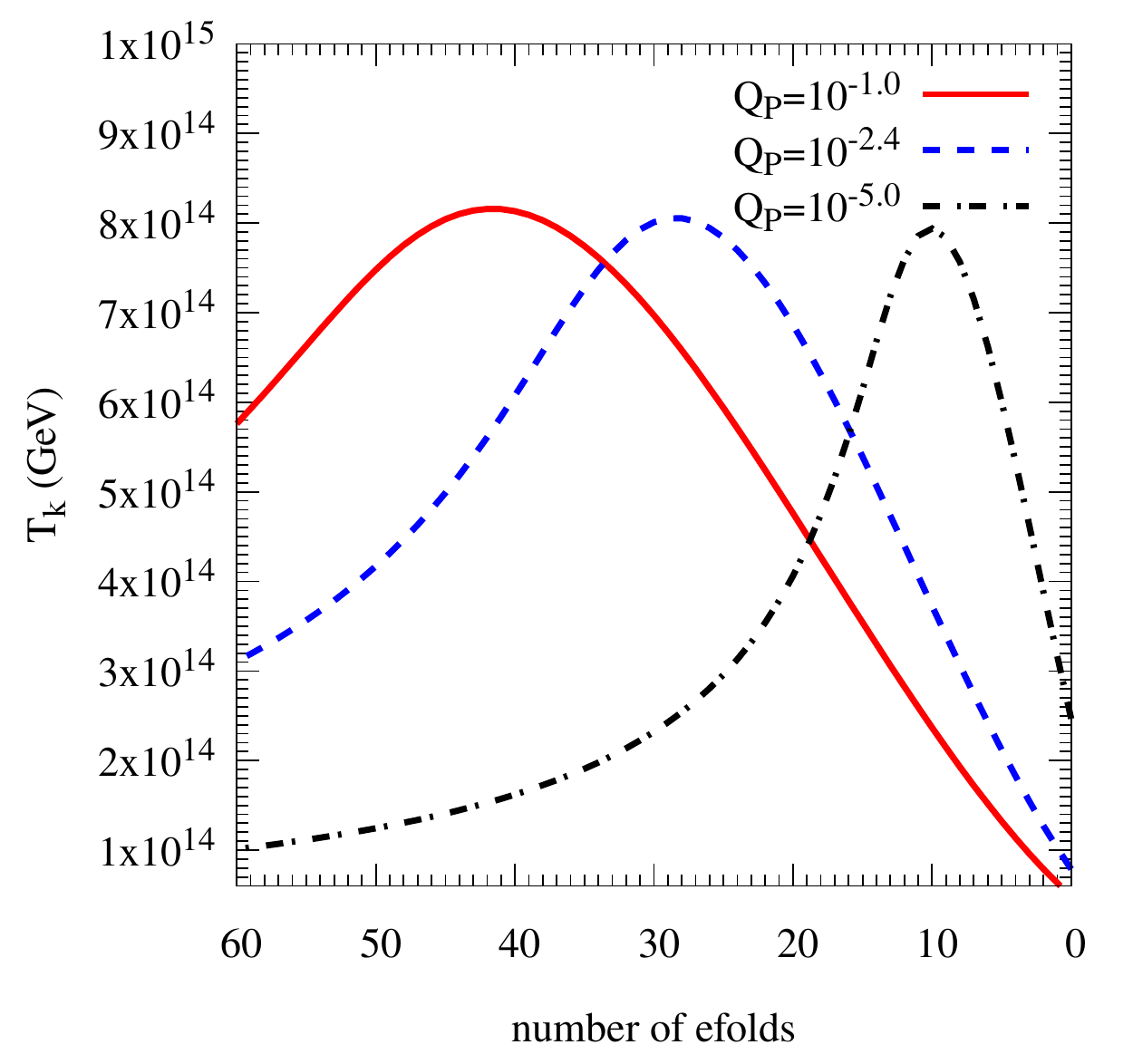}
  \vspace{-0.8cm}
  \caption{Plots showing  $T_k/H_k$ vs $\ln(k/k_P)$ 
  and $T_k $ vs $\ln(k/k_P)$ for different values of $Q_P$.
  Note that $T>H$ during inflation is valid only for $Q_P>10^{-5}$.
 }
  \label{Tvsx}
\end{figure}

As mentioned earlier, the temperature is defined only for $T_k>H_k$. 
In Figure (\ref{Tvsx}), we plot $T_k/H_k$ as a function of the number of e-foldings $N$
for different values of $Q_P$, and $N_P=50$ and fixed $A_s$.  
We find that $T_k>H_k$ holds for $Q_P>10^{-5.0}$. One also notices that $T_k/H_k$ increases with time 
till the end of inflation.  We also plot $T_k$ as a function of $N$.
(We mention in passing that the relation between $Q_P$ and $T/H$ in Sec. 4 of 
Ref. \cite{Imp} has a typographical error.)

\subsection{Correlation between $\lambda$ and $Q_P$}
\label{PofkP}

We note that for $Q_P\ll1$ sec. \textsection \ref{powerspectrum} implies that
\begin{eqnarray}
A_s = P_{\cal R}(k_P)&\sim \lambda \left(\frac{\phi_P}{\mpl}\right)^6  SB \; , \\
&\sim \lambda \left(\frac{\lambda C_\phi^4}{Q_P}\right)^\frac{6}{10} SB\,,
\end{eqnarray}
where $SB$ refers to the square bracket in Eq. \eqref{Pkfull}.
The dependence of $\lambda C_\phi^4$ on $Q_P$ is shown in Figure (\ref{fig:cphivsQP}).
For $N_P=50$ and fixed $A_s$ we find how $SB$ varies with $Q_P$, and this is
also shown in Figure (\ref{fig:cphivsQP}).
Numerical fitting tells us that 
\begin{align}
\lambda C_\phi^4 &\sim Q_P^{0.60} \; ,
\label{lambdaCphi4QPreln} \\
SB &\sim Q_P^{0.35} \; ,
\label{SBQPreln}
\end{align}
and so
\vspace{-0.5cm}
\begin{align}
P_{\cal R}(k_P)&\sim  \lambda \left(\frac{Q_P^{0.60}}{Q_P}\right)^\frac{6}{10} Q_P^{0.35} \; ,\nonumber \\
&\sim  \lambda Q_P^{0.11} \; ,
\label{PRkP}
\end{align}
which implies $\lambda\sim Q_P^{-0.11}$.
(The factorisation of the r.h.s. of Eq. \eqref{PRkP} 
into powers of $\lambda$ and $Q_P$ is not unique.  For example,
$SB$ contains $T_P/H_P$ which has a factor of $\lambda^{-\frac14}$. 
One could try to
extract it out explicitly
or subsume
it
in the behaviour of $SB$ as 
a function of $Q_P$, as we have done.  Either way, 
the relation between $\lambda$ and $Q_P$
will be the same and as obtained here.)

In Sec. \textsection \ref{couplings} we had discussed the relation between $\lambda$ and $Q_P$ for fixed $A_s$ and $N_P$ and found it
to be $\lambda = Q_P^{-0.10}\times 10^{-14.0}$.
Obviously, the fitting function between $\lambda$ and $Q_P$ must be the same no matter how it is found, and we find that the $\lambda-Q_P$ relations found 
earlier and in this subsection
are mutually consistent,  confirming our analysis. 
Thus, we expect that when we perform the {\tt CosmoMC} analysis,
we should obtain for the joint distribution of $\lambda$ and $Q_P$
 a correlation between the two characterised by $\lambda \sim Q_P^{-0.1}$ for $N_P=50$. 
 Repeating the above analysis for $N_P=60$ also gives $\lambda \sim Q_P^{-0.1}$ and we expect {\tt CosmoMC} to give a similar result. 

\section{Analysis of the effects of warm inflation parameters on the CMB angular power spectrum}

\subsection{$\lambda$ and $Q_P$}

\begin{figure}[h]
  \includegraphics[width=0.5\textwidth]{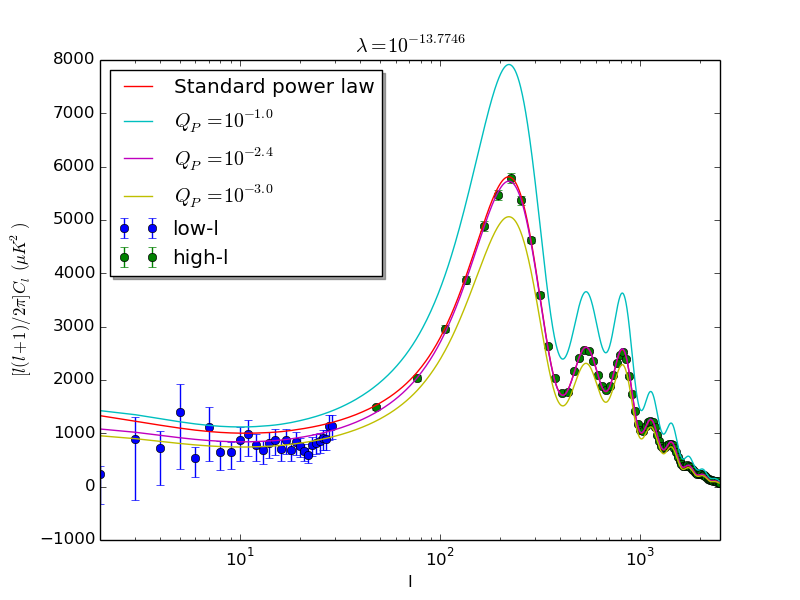}
  \includegraphics[width=0.5\textwidth]{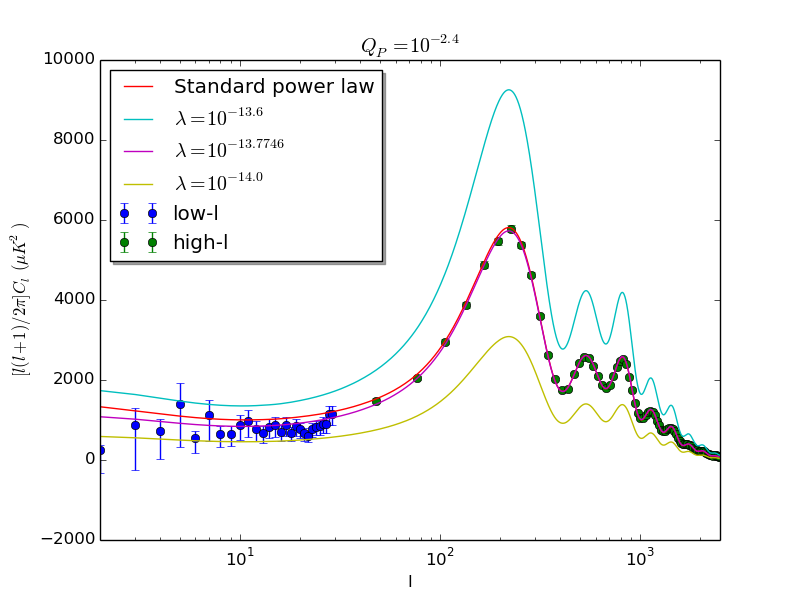} \vspace{-0.7cm}
\caption{The plot on the left shows the TT angular power spectrum of the CMB for various values of $Q_P$, for $\lambda$ fixed at $10^{-13.7746}$ and $N_P$ fixed at $50$.
The plot on the right shows the TT angular power spectrum of CMB for various values of $\lambda$, for $Q_P$ fixed at $10^{-2.4}$ and $N_P=50$.
For both plots, the best fit corresponds to $Q_P=10^{-2.4}$ and 
$\lambda=10^{-13.7746}$, which overlaps with the $C_{\ell}$'s obtained
with the standard power law spectrum with $n_s=0.96$. 
In the plots, low $\ell$ refers to the range $\ell = 2 - 49$ while high $\ell$ refers to the range $\ell = 50 - 2500$ (see Ref. \cite{Ade:2013kta} for details).
} 
\vspace{-0.2cm}
\label{fig:Clfixedlambda}
\end{figure}

 In Figure (\ref{fig:Clfixedlambda})
we plot the $TT$ angular power spectrum of CMB for fixed $\lambda$ 
 while varying $Q_P$, and for fixed $Q_P$ while varying $\lambda$, respectively keeping $N_P=50$ fixed. We obtain the plots using {\tt CAMB}. We see that the angular power spectrum is sensitive to even a slight variation in $\lambda$.  However it varies considerably only with order of magnitude changes in $Q_P$.
 This behaviour is also suggested by Eq. \eqref{PRkP}.
 
\subsection{Effect of the {\it coth}  term, and a comparison with cold inflation}
\label{cothsection}

A Bose-Einstein distribution for inflatons has been considered in the context 
of cold inflation too with a similar $\coth[H_k/(2T_k)]$ term \cite{Bhattacharya:2005wn}.  In that case, it is presumed that the inflaton was in thermal
equilibrium at some early time prior to inflation, possibly at the Planck time, and subsequently its interactions froze out.  Therefore one has a frozen
distribution of inflatons at the beginning of inflation.  Since this distribution
dilutes with the expansion of the Universe, and hence during inflation, the 
effect of the thermal distribution is larger for larger scales.  In fact, it enhances
power on larger scales and if the power spectrum is pivoted at a standard pivot scale then,
as seen in Fig. (1) of Ref. \cite{Bhattacharya:2005wn}, there is excess power on large scales  which requires inflation to last 
7-32 more e-foldings than the standard requirement for  GUT scale to electoweak scale inflation
so as 
to dilute the effect of the inflaton distribution.

\begin{figure}[h]
	\centering
	\includegraphics[width=9cm]{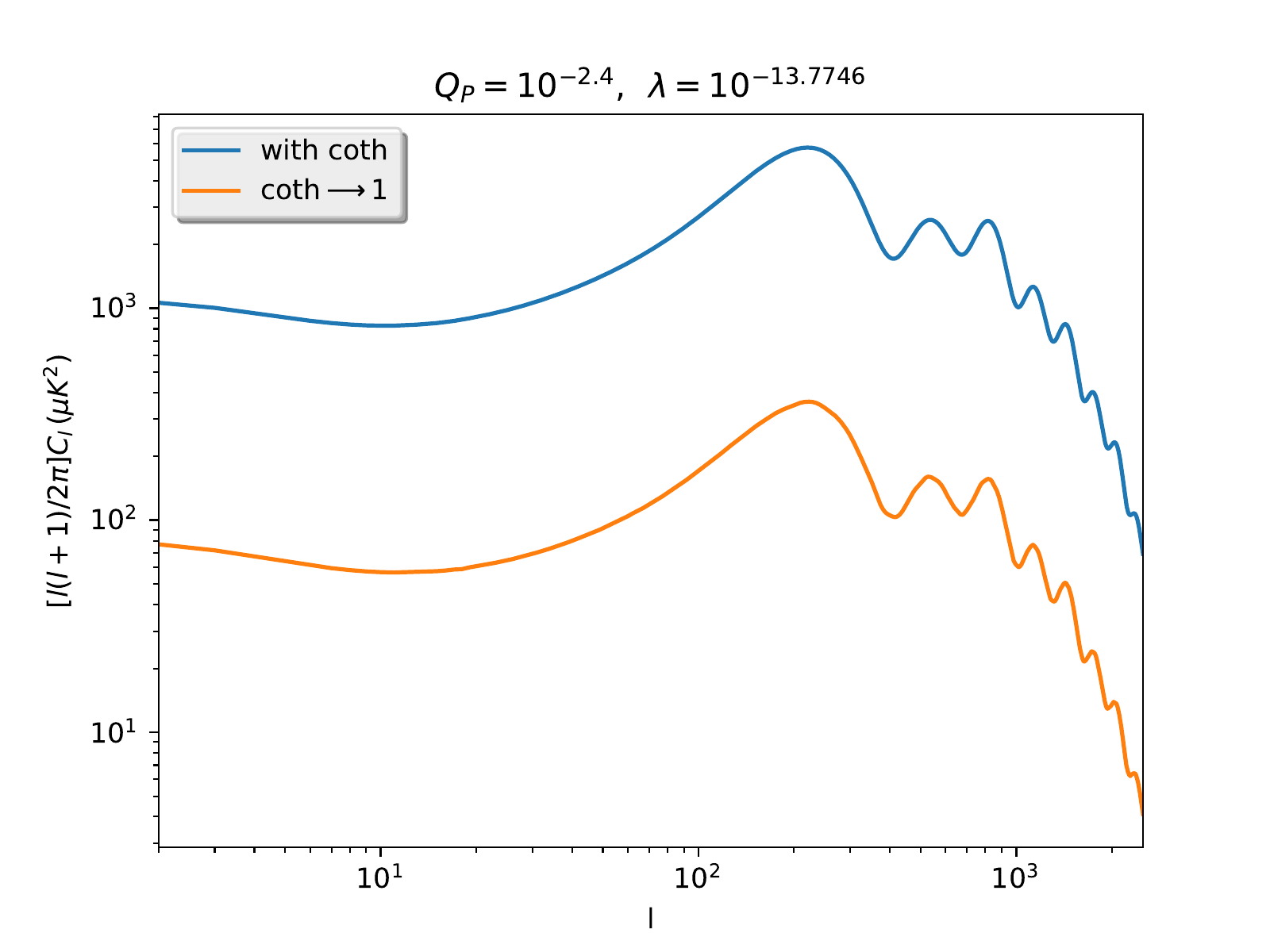} 
	\vspace{-0.5cm}
	\caption{
		Angular power spectra for warm inflation for the same values of $Q_P$ and 
		$\lambda$ are plotted using {\tt CAMB}, with the coth term retained in the primordial power spectrum for the upper blue curve and with the coth term set to 1 for the lower orange curve.
		One sees that the effect of the coth term is to enhance the angular power spectrum, and it is almost
		uniform across angular scales.  
		The parameters $Q_P$ and $\lambda$ are appropriately chosen 
		such that at the pivot scale the primordial power spectrum associated with the blue curve is normalised.
	} 
	\label{fig:cotheffect}
\end{figure}

For warm inflation where one considers a thermal distribution of inflatons
the distribution is replenished by dissipation and the temperature is approximately constant when cosmologically relevant $k$ modes leave the horizon.  
For $N_P=50$ and fixed $A_s$, and $Q_P=10^{-2.4}$,
we find that $\coth[H_k/(2T_k)]$ varies from 13 to 26
as $x$ varies from $-7$ to 
$3$.
The ratio of the square bracket SB in the primordial power spectrum with the coth
and with the coth set to 1 varies by a factor of 13 to 17, which is also reflected in the $C_\ell$'s in 
Figure (\ref{fig:cotheffect}), which shows the angular power spectrum for 
the same value of $Q_P$ and $\lambda$ with the coth term retained and the coth term
set to 1.

Because the coth term shows only some variation with $k$, it effectively
readjusts the 
normalisation of the prefactor in the power
spectrum, i.e., it lowers $H$.
For $N_P=50$ and fixed $A_s$, and 
$Q_P=10^{-2.4}$ for warm inflation, we have calculated the parameters of warm inflation 
with the coth term retained, 
with the coth term set to 1, 
and for cold inflation.  
The $\lambda, \phi_P$ and
$H_P$ are  $(1.68 \times 10^{-14}, 3.01 M_{Pl}, 4.09 \times 10^{13} {\rm GeV}),
(3.09 \times10^{-13}, 3.01 M_{Pl}, 1.76 \times 10^{14} {\rm GeV})$ 
and  
$(6.14 \times 10^{-14}, 4.03 M_{Pl}, 1.40 \times 10^{14} {\rm GeV})$ 
respectively for the three cases.
One finds that $H_P$ for the warm inflation case with the coth term retained
is the lowest. Lower $H_P$ lowers the tensor power spectrum
and hence the tensor to scalar ratio, $r$, thereby making the quartic potential warm inflation 
scenario with the coth term retained more compatible with Planck data.
Quartic warm inflation with $n(k)=0$, and  quartic cold inflation, give too large values of
$r$ and are incompatible with the data, as also shown in Ref. \cite{Imp}.

\subsubsection{Recovering cold inflation}

For $Q_P<10^{-10}$  the value of $\lambda$ does not change with $Q_P$ for fixed
$N_P$ and $A_s$ indicating that perhaps one has achieved the cold inflation limit.
Indeed we find that the value of $\lambda$ and $n_s$ for $Q_P=10^{-10}$
are $5.96\times10^{-14}$ and 0.94 and the corresponding numbers for cold
inflation are very similar at $6.14\times10^{-14}$ and 0.94.
Note that for such small $Q_P$, $T/H$ is small and the coth term reduces to 1 giving
a power spectrum corresponding to $n_k\approx 0$, as for standard cold inflation, which was also noted in 
Ref. \cite{Imp}. 

\section{Comparison with the existing literature}

An analysis of the power spectrum and its dependence on the warm inflation
model parameters is also given in Sec. III of 
Ref. \cite{Mar-Berera-IJMP}.
\footnote{
The form of the power spectrum in Eq. (30) of Ref. \cite{Mar-Berera-IJMP} reduces to
that which we consider below in the weak dissipative limit $(Q\ll1)$ with a thermal distribution for the inflaton quanta.
There is a discrepancy of a factor of 2,
and terms that reduce to 1 in the small $Q$ limit.}  
Their Table 3  implies that $\lambda\sim C_\phi^{-1}\sim Q_P^{-0.33}$ which differs 
somewhat from our
result in Sec. \ref{couplings} that $\lambda \sim
C_\phi^{-0.57}\sim Q_p^{-0.1}$ for $N_P$
equal to 50, which is  corroborated by our {\tt CosmoMC} analysis in 
Sec. \textsection \ref{cosmomc}.

\begin{figure}[h]
	\centering
\includegraphics[width=9cm]{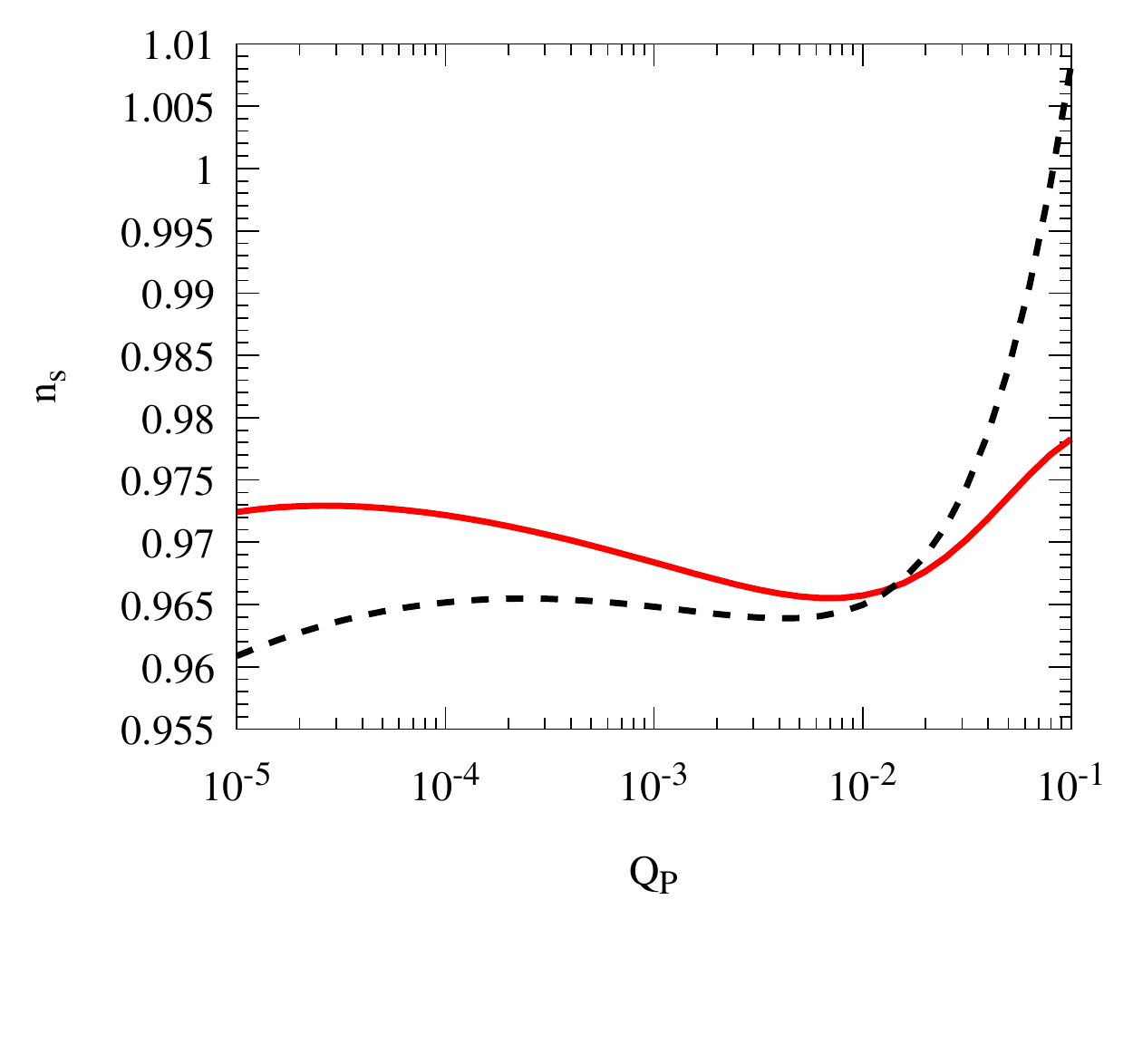}
	\vspace{-1.5cm}
	\caption{ $n_{s}$ vs $Q_P$ from our expression in Eq. \eqref{nsfull}  (red solid line), and from that of Ref. \cite{CosmoFluids} in Eq. \eqref{nsCF} (black dashed line) for $N_P=50$ and fixed $A_s$. 
	}
	\label{nscomparison}
\end{figure}

In Eq. (5.23) of Ref. \cite{CosmoFluids}, the scalar spectral index for $Q_P\ll1$
 is given by the expression
 \begin{equation}
n_s= 1- 3\epsilon_1 + \frac{1}{(1+2 n_P+ \Delta_Q)}  4   n_P \  \epsilon_1 + \frac{1}{(1+2 n_P+ \Delta_Q)}7 \epsilon_1 \Delta_Q \,,
\label{nsCF}
\end{equation}
where $$\Delta_Q=2\pi Q_P \frac{T_P}{H_P}. $$
We have plotted our expression for
$n_s$ in Eq. \eqref{nsfull} , and that in Eq. \eqref{nsCF}, 
as a function of $Q_P$ in Figure (\ref{nscomparison}) for $Q_P\le 0.1$.
One sees significant differences in the plots\footnote{Our analysis is for $N_P=50$ and fixed $A_s$. However, if we do not fix $A_s$ but take a fixed $\lambda$ (like in Figs. 1 and 3 of Ref. \cite{CosmoFluids}), one sees  differences in the $n_s$ plots obtained using Eqs. \eqref{nsfull} and \eqref{nsCF} similar to that in Figure (\ref{nscomparison}).}.
The expression  in  Ref. \cite{CosmoFluids} is obtained using
certain approximations and hence does not include certain factors and terms.
First of all, because of the coth$\left(\frac{H}{2T}\right) \sim \frac{T}{2H}$ for $H\ll2T$ approximation \footnote{In Sec. 5 of Ref. \cite{CosmoFluids}
	the argument of the coth term and the approximation for
	coth have typographical errors.
}, a factor $\frac{e^{H_P/T_P}}{e^{H_P/T_P}-1} \frac{H_P}{T_P}$ was missed by the authors.
 This factor varies from 1 to 1.4 as $Q_P$ goes from $10^{-1}$ to $10^{-5}$. 
Secondly, Eq. \eqref{nsCF} is valid only in the $Q_P\ll 1$ approximation which is indicated by the statement just above their Eq. (5.23). If the approximation $Q_P\ll 1$ is not considered and $1+Q_P$ is retained in the derivation, an extra term $\frac{10 \epsilon_1 Q_P}{(1+7Q_P)(1-\epsilon_1)}$ is obtained as in our Eq. \eqref{nsfull}. This extra term is significant, $\Delta n_s \ge 0.005$, for $Q_P > 10^{-1.8}$. 
 Thirdly, in Ref. \cite{CosmoFluids} the spectral index is defined as $d \ln P_{\cal R}/dN$ instead of $d \ln P_{\cal R}/d\ln(k/k_P)$, because of which another factor
 	of $dN/dx=-(1-\epsilon_1)^{-1}$, as given in Eq. (\ref{dNdx}), is ignored
 \footnote{In Ref. \cite{CosmoFluids}, $dN$ is defined as the negative of ours and hence there is no sign discrepancy.}. The magnitude of this overall factor
	varies from 1.05 to 1.02 for $Q_P$ between $10^{-1}$ to $10^{-5}$.  
In all, the combined effect of these three factors is seen in Figure (\ref{nscomparison}).
In the $Q_P \ll
1$ limit, 
and if we drop the factor $\frac{e^{H_P/T_P}}{e^{H_P/T_P}-1} \frac{H_P}{T_P}$,
our expression for the spectral index agrees with that in Ref. \cite{CosmoFluids}
 which implies that Eq. \eqref{nsCF} is equivalent to Eq. \eqref{nsfull} combined with the above approximations.

\section{Constraining the parameters of the warm quartic chaotic inflation
model with the  full Planck 2015 data}  
\label{cosmomc}

As we saw in Section \textsection 
\ref{powerspectrum}, for warm inflation driven by a single scalar field with a quartic self-interaction potential, for a given $N_P$ the parameters which determine the underlying microphysics can be chosen to be
$Q_P$ and $\lambda$. We therefore obtain the joint and marginal distributions of these parameters and constrain their values.

It is well known that in cold inflation driven by a quartic potential, the self coupling $\lambda$ is quite small (${\cal O}(10^{-14}))$.  From our preliminary analysis with {\tt CAMB} we have found that
$\lambda$ for warm inflation is similarly small.  Hence
we work with $-\log_{10} \lambda$ rather than $\lambda$
as one of the variables in performing our analysis.
In the weak dissipative regime  $Q_P \ll 1$
and we use  $-\log_{10} Q_P$ as another variable. We can now attempt to constrain $-\log_{10} \lambda$ and $-\log_{10} Q_P$ for the two choices of $N_P$ i.e. $N_P = 50$ and $N_P = 60$. 

For $N_P=50$ and 60, we use $10^{-14}$ to $10^{-13}$ and  
$10^{-14.5}$ to $10^{-13.5}$
respectively as the range of values for
$\lambda$.  The angular power spectrum is sensitive to $\lambda$ and we have to
choose a narrow range.
As discussed in Section 
\textsection \ref{sec:TgtH}
for the values of $Q_P$ smaller than $10^{-5}$, one has $T < H$ and so this regime can not be thought of as a realisation of warm inflation. Keeping this in mind, we 
would like to 
consider $Q_P$ between $10^{-5}$ and $10^{-1}$.  However for numerical reasons we consider
$-\log_{10} Q_p$ between 5.4 and 1 for both $N_P=50$ and 60.

In  the Planck 2015 likelihood analysis  \cite{planck2015:likelihood} the CMB  angular power spectra 
at low-l are computed following a Gibbs sampling approach called {\tt COMMANDER}. At high-l, the
TT, TE and EE   power spectra are computed by using the code {\tt Plik} which computes ``Pseudo-Cls'' by
cross-correlating the data from different frequency maps. {\tt Plik} has a large number nuisance
parameters, representing the foreground, calibration, etc., for computing likelihood.
Although the parameters of {\tt Plik} also can be estimated from the Planck 2015 data in a
{\tt CosmoMC} analysis  we have decided not to do that and fixed their values to be the best
fit values for the standard $\Lambda$CDM baseline model as given by the Planck collaboration.
We carry out our MCMC search in a six dimensional parameter space
with the standard parameters $\Omega_bh^2, \Omega_ch^2, \theta, \tau$, and our
parameters $-\log_{10}\lambda$ and  $-\log_{10}Q_p$.  $\Omega_bh^2, \Omega_ch^2, \theta$ and $\tau$ represent 
the baryon density, cold dark matter density, 
the ratio of the size of the sound horizon at decoupling ($r_{dec}$) and the angular diameter
distance at decoupling ($D_A$), and the
Thomson scattering optical depth due to reionization respectively.
The priors for the standard parameters are found by comparing the CMB angular power spectra for a 
set of trial values with the observational data points using {\tt CAMB}. 
We consider flat priors for all these parameters.
We use the November 2016 version of {\tt CAMB}
and the July 2015 version of {\tt CosmoMC} and set the flags,  compute\_tensor=T, CMB\_lensing=T, and
use\_nonlinear\_lensing=F. We use as the pivot scale $k_{P}=0.05 \text{Mpc}^{-1}$, set the number of massless
neutrino species to its value in the Standard Model i.e. $n_{\nu}=3.046$ and similarly set the value of 
the Helium fraction $Y_{He}=0.24$ in our analysis. 

\begin{figure}[h]
\begin{center}
\includegraphics[width=\textwidth]{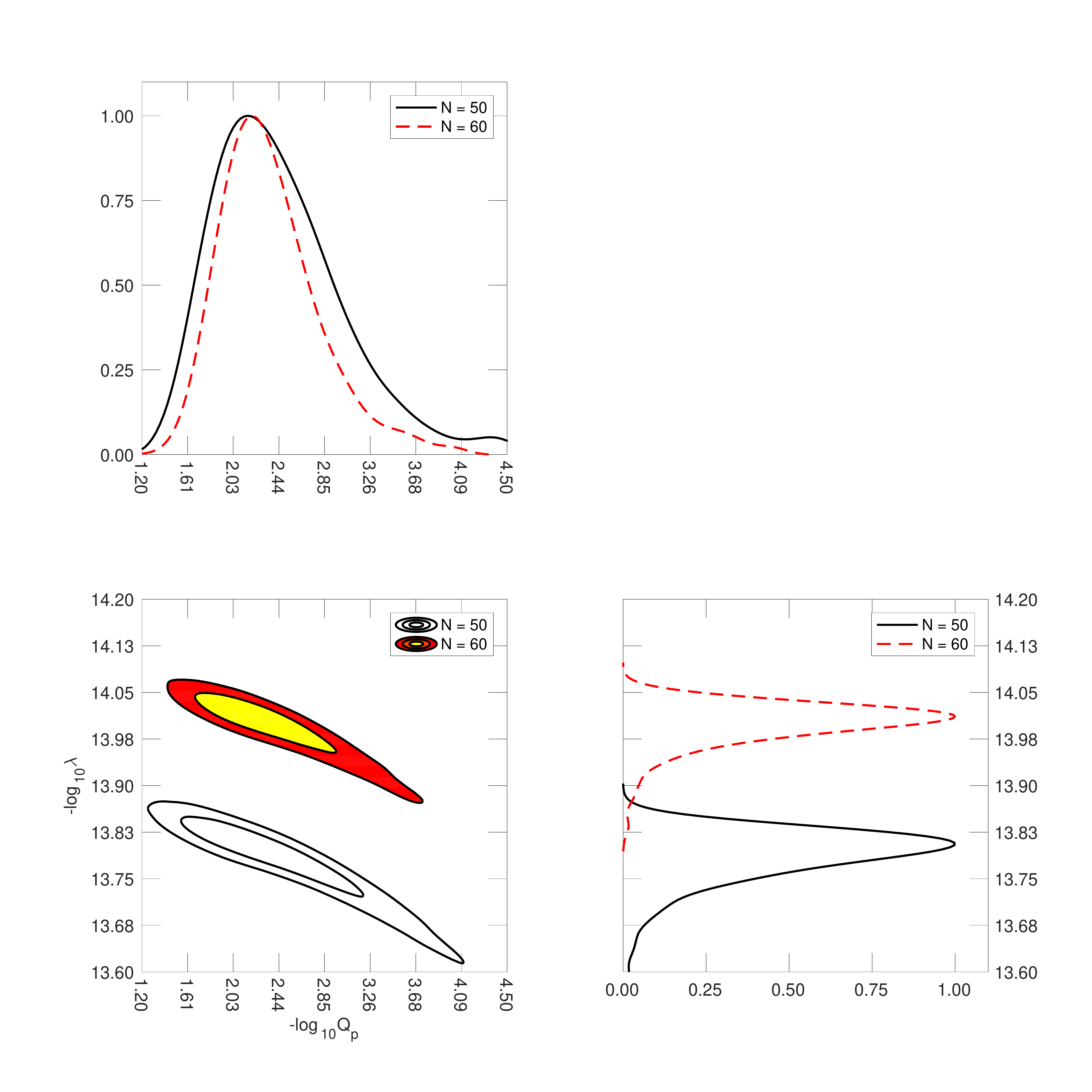}
\vspace{-1.5cm}
\caption{The joint probability distribution and the marginalized distributions for $\lambda$ and $Q_p$. 
For the marginalized distributions  the black solid lines 
are for $N_P=50$ and the red dashed lines are for $N_P=60$.
In the contour plots of the joint probability distributions the lower contours are for $N_P=50$ and the upper ones are for $N_P=60$.
}
\label{jointprob}
\end{center}
\end{figure}

We present the mean and 68 \% limits 
for the six parameters we consider in  Table \ref{tab:bestfit}.
The mean values of $\lambda$ for $N_P=50$ and 60 are
$1.6\times 10^{-14}$ and
$1.0\times10^{-14}$
respectively,
and of $Q_P$ are 
$3.7\times10^{-3}$ and
$4.4\times10^{-3}$
respectively.
We also show the marginalized probability distributions 
and the joint probability distribution of these parameters 
in Figure (\ref{jointprob}). 
We find that both the parameters of the quartic warm inflation model that we consider,
$\lambda$ and $Q_P$, are well constrained by the Planck 2015 temperature and polarization 
data. 
The shape of the joint probability distribution of 
$-\log_{10}\lambda$ and  $-\log_{10}Q_P$ in Figure (\ref{jointprob})
is best understood by comparing with the 
$\log_{10}\lambda$ vs  $\log_{10}Q_P$ plot in Figure (\ref{fig:cphivsQP}), 
and is consistent with the empirical relation $\lambda \sim Q_p^{-0.1}$  
for  $N_P = 50$ and 60 obtained in Section \textsection \ref{PofkP}.

\begin{table}[ht]
 \begin{center}
 \begin{tabular}{|c|c|c|c|c|}\hline \hline
 { S. No.} & { Parameter} & { Priors}  & { $N_P=50$} & { $N_P=60$}  \\ \hline \hline
            1 & $\Omega_bh^2$ & (0.005,0.1)  & $       0.02220 \pm       0.00013 $&$       0.02226 \pm       0.00013 $ \\ \hline
            2 & $\Omega_ch^2$ & (0.001,0.99) & $        0.1191 \pm        0.0010$&$         0.1177 \pm        0.0009 $ \\ \hline
           3 & $100\theta$   & (0.50,10.0)  &  $       1.04089 \pm       0.00029$&$        1.04105 \pm       0.00029 $ \\ \hline
         4 & $-\log_{10}\lambda$ &
         (13.0,14.0),(13.5,14.5)&         
         $        13.783 \pm         0.051 $&$        13.998 \pm         0.037 $ \\ \hline
          5 & $-\log_{10}Q_P$ &  (1.0,5.4)   & $        2.4358 \pm        0.5856 $&$        2.3585 \pm        0.4495 $ \\ \hline
                   6 & $\tau$ &  (0.1,0.8)   & $         0.065 \pm         0.011 $&$         0.075 \pm         0.011$  \\ \hline
\end{tabular}
\caption{
From the Planck 2015 TT, TE, EE + low P + lensing  data mean and 68\% limits for the
 four standard and two extra parameters of the quartic model of warm inflation 
 for 50 
 and 60  
 number of e-foldings is displayed.
 The priors for various parameters are also included.  (For $-\log_{10}\lambda$ separate priors
 for $N_P=50$ and 60 are indicated.) 
 }
\label{tab:bestfit}
\end{center}
\end{table}

A {\tt CosmoMC}  analysis of warm inflation for different potentials was carried out in Ref. \cite{Benetti:2016jhf}.  
In that analysis the {\tt CosmoMC} parameters are $\Omega_c h^2$, $\Omega_b h^2$, $\theta$, $\tau$ and $Q_P$, for  fixed $N_P$.  The normalization of the power spectrum at the pivot scale is also taken to be fixed. In the power spectrum, the  authors parametrize $Q_k$, $T_k/H_k$ and the prefactor in the 
power spectrum, $P_0(k)=[H_k^2/(2\pi\dot\phi_k)]^2$, as $a_i k^{b_i + c_i \ln(k) }$ with $i$ referring to $Q_k$, $T_k/H_k$ and $P_0(k)$ 
\footnote{R. Ramos, private communication.}.
For each value of $Q_P$,  and fixed $P_{\cal R}(k_P)$ and $N_P$, their background equations
[Eqs. (2.1-2.3, 2.9-2.15)] are used to (uniquely) determine $a_i, b_i$ and $c_i$.  With the values of these coefficients $P_\mathcal{R}(k)$ is known, which is used by {\tt CosmoMC}.
{\tt CosmoMC} then provides a likelihood curve for $Q_P$, as shown in their Fig. 8.  The algorithm BOBYQA (Bound Optimization BY Quadratic Approximation) is used to find the best fit value. Furthermore they obtain $n_s, n_s'$ and $n_s^{''}$  at $k_P$
using their Eqs. (2.23-2.25).
Planck 2015 results suggest a lower value of the tensor to scalar $r$, and a 
positive value of the running of the running parameter 
$n_s^{''}$ (albeit of small statistical significance) and these cannot be explained 
in the standard cold inflation scenario  driven by a quartic potential. 
In Ref. \cite{Benetti:2016jhf} it has been shown that different
warm inflation models can 
explain these features, though not quartic warm inflation.

In the present work, instead of restricting our attention to  finding only the best fit value of $Q_P$ (and then the other derived parameters), we vary both $Q_P$ and $\lambda$.  This allows us to get the joint probability distribution for the two parameters which describe the underlying microphysics of the
quartic chaotic warm inflation model.
Furthermore, we do the full {\tt CosmoMC} and not just BOBYQA.
This allows us to get the mean values and standard deviations.  In algorithms such as those
followed in {\tt CosmoMC} the range of preferred values, represented by
the mean and standard deviation, carry more information than the best
fit value.

\section{Discussion and Conclusions}

The standard cold inflation is based on the simple picture in which the inflaton does not experience any dissipative dynamics as it slowly rolls down during the inflationary phase. On the other hand, warm inflation provides an example of an alternative picture, in which the dissipative dynamics plays a major role even during slow roll.
The next generation CMB experiments will provide unprecedentedly stringent constraints on the various parameters characterizing the primordial power spectrum of large field inflationary scenarios \cite{Abazajian:2016yjj,Munoz:2016owz}. Given this, it is important to understand how these constraints are to be interpreted, i.e. what will these constraints teach us about the epoch of cosmic inflation. Keeping this future prospect in mind, one must first best understand how the current available data constrains the possibility of warm inflation.

In this work, we studied warm inflation driven by a  scalar field with a quartic potential in the weak dissipative regime i.e. in the regime in which the inflaton decay rate is small as compared to the Hubble parameter when the modes of cosmological interest cross the horizon during inflation. We focused on constraining all aspects of this specific scenario using the most recent CMB observational data. 
In particular we found the joint probability distribution of $\lambda$, the self coupling of the inflaton, and $Q_P$, the ratio of the inflaton decay width and (three times) the Hubble parameter during inflation, evaluated at the pivot scale (0.05 ${\rm Mpc}^{-1}$).  We also found the mean 
 and standard deviation values of these parameterss.  $Q_P$ is also related to $C_\phi$, a parameter
related to the inflaton coupling to other fields and the number of other fields in the warm inflation
model.

From the marginalised distributions of the parameters of the model we find that 
the preferred range of values of $\lambda$ is
$1.5\times10^{-14} $  to  $1.9\times10^{-14}$  with a mean value of $1.6\times 10^{-14}$
for the number of e-foldings of inflation after the pivot scale 
crosses the horizon during inflation, $N_P$, equal to 50, and
the preferred range of values for $N_P$ equal to 60 is
$9.2\times10^{-15}$   to   $1.1\times10^{-14}$ with a mean value of $1.0\times10^{-14}$.
The preferred range of values of $Q_P$ is 
$9.5\times10^{-4}$   to $1.4\times10^{-2}$  with a mean value of $3.7\times10^{-3}$ for
$N_P$ equal to 50, and
the preferred range of values for $N_P$ equal to 60 is
$1.6\times10^{-3}$  to $1.2\times10^{-2}$ with a mean value of $4.4\times10^{-3}$.   
The values of $C_\phi$ corresponding to the mean values of $Q_P$ are 
$8.7\times10^6 \,(1.2\times10^7)$ for $N_P$ equal to 50 (60), 
from the analysis in 
Sec. \textsection \ref{couplings}.

From the joint probability distribution obtained
using {\tt CosmoMC}, as well us from the analysis of the power spectrum, normalised at the pivot
scale, in Secs. \ref{couplings} and \ref{PofkP}, we obtain $\lambda \sim Q_P^{-0.1}$ for both $N_P=50$ and 60,  
a potentially useful result for warm inflation model building. 
 
 {\bf Acknowledgements:} It is a pleasure to acknowledge discussions with Arjun Berera, Mar
 Bastero-Gil and Rudnei Ramos about details of their work on the warm inflationary
 scenario. J.P. would like to thank the Navajbai Ratan Tata Trust (NRTT) for financial support.
 Numerical work for the present study was done on the IUCAA HPC facility.
 
 {\bf Note:} 
 While this manuscript was being prepared, Ref. \cite{Bastero-Gil:2017wwl} appeared on the arXiv. 
 They too consider a {\tt CosmoMC} analysis of warm inflation. They consider an inflaton dissipation rate proportional to the temperature $T$, while above we consider $\Upsilon \sim T^3$.

\bibliographystyle{JHEP}

\begin{thebibliography}{99}
 
\bibitem{Starobinsky:1980te} 
  A.~A.~Starobinsky,
  ``A New Type of Isotropic Cosmological Models Without Singularity,''
  Phys.\ Lett.\  {\bf 91B}, 99 (1980).
 
  
\bibitem{Kazanas:1980tx} 
  D.~Kazanas,
  ``Dynamics of the Universe and Spontaneous Symmetry Breaking,''
  Astrophys.\ J.\  {\bf 241}, L59 (1980).
  
\bibitem{Guth:1980zm} 
  A.~H.~Guth,
  ``The Inflationary Universe: A Possible Solution to the Horizon and Flatness Problems,''
  Phys.\ Rev.\ D {\bf 23}, 347 (1981).
  
  
\bibitem{Linde:1981mu} 
  A.~D.~Linde,
  ``A New Inflationary Universe Scenario: A Possible Solution of the Horizon, Flatness, Homogeneity, Isotropy and Primordial Monopole Problems,''
  Phys.\ Lett.\  {\bf 108B}, 389 (1982).
  
\bibitem{Albrecht:1982wi} 
  A.~Albrecht and P.~J.~Steinhardt,
  ``Cosmology for Grand Unified Theories with Radiatively Induced Symmetry Breaking,''
  Phys.\ Rev.\ Lett.\  {\bf 48}, 1220 (1982).
  
\bibitem{Starobinsky:1979ty} 
  A.~A.~Starobinsky,
  ``Spectrum of relict gravitational radiation and the early state of the universe,''
  JETP Lett.\  {\bf 30}, 682 (1979)
  [Pisma Zh.\ Eksp.\ Teor.\ Fiz.\  {\bf 30}, 719 (1979)].
  
\bibitem{Hawking:1982cz} 
  S.~W.~Hawking,
  ``The Development of Irregularities in a Single Bubble Inflationary Universe,''
  Phys.\ Lett.\  {\bf 115B}, 295 (1982).
  
\bibitem{Starobinsky:1982ee} 
  A.~A.~Starobinsky,
  ``Dynamics of Phase Transition in the New Inflationary Universe Scenario and Generation of Perturbations,''
  Phys.\ Lett.\  {\bf 117B}, 175 (1982).
  
\bibitem{Guth:1982ec} 
  A.~H.~Guth and S.~Y.~Pi,
  ``Fluctuations in the New Inflationary Universe,''
  Phys.\ Rev.\ Lett.\  {\bf 49}, 1110 (1982).

 

  
\bibitem{Berera:1995wh} 
  A.~Berera and L.~Z.~Fang,
    ``Thermally induced density perturbations in the inflation era,''
      Phys.\ Rev.\ Lett.\  {\bf 74}, 1912 (1995)
      [astro-ph/9501024].
  
\bibitem{Berera:1995ie} 
  A.~Berera,
    ``Warm inflation,''
      Phys.\ Rev.\ Lett.\  {\bf 75}, 3218 (1995)
      [astro-ph/9509049].
	  
	  \bibitem{Berera:1999ws} 
	    A.~Berera,
	    ``Warm inflation at arbitrary adiabaticity: A Model, an existence proof for inflationary dynamics in quantum field theory,''
	    Nucl.\ Phys.\ B {\bf 585}, 666 (2000)
	    [hep-ph/9904409].
	    
	  \bibitem{LopezNacir:2011kk} 
	    D.~Lopez Nacir, R.~A.~Porto, L.~Senatore and M.~Zaldarriaga,
	    ``Dissipative effects in the Effective Field Theory of Inflation,''
	    JHEP {\bf 1201}, 075 (2012)
	    [arXiv:1109.4192 [hep-th]].

\bibitem{Berera:1998px} 
  A.~Berera, M.~Gleiser and R.~O.~Ramos,
  ``A First principles warm inflation model that solves the cosmological horizon / flatness problems,''
  Phys.\ Rev.\ Lett.\  {\bf 83}, 264 (1999)
  [hep-ph/9809583].

\bibitem{Bastero-Gil:2015nja} 
  M.~Bastero-Gil, A.~Berera and N.~Kronberg,
  ``Exploring the Parameter Space of Warm-Inflation Models,''
  JCAP {\bf 1512},  046 (2015)
  [arXiv:1509.07604 [hep-ph]].

\bibitem{Mishra:2011vh} 
  H.~Mishra, S.~Mohanty and A.~Nautiyal,
  ``Warm natural inflation,''
  Phys.\ Lett.\ B {\bf 710}, 245 (2012)
  [arXiv:1106.3039 [hep-ph]].

\bibitem{Bastero-Gil:2016qru} 
  M.~Bastero-Gil, A.~Berera, R.~O.~Ramos and J.~G.~Rosa,
  ``Warm Little Inflaton,''
  Phys.\ Rev.\ Lett.\  {\bf 117},  151301 (2016)
  [arXiv:1604.08838 [hep-ph]].

\bibitem{cosmomc}
  A.~Lewis and S.~Bridle,
  ``Cosmological parameters from CMB and other data: A Monte Carlo approach,''
  Phys.\ Rev.\ D {\bf 66}, 103511 (2002)
  [astro-ph/0205436].

  \bibitem{Moss:2006gt} 
    I.~G.~Moss and C.~Xiong,
    ``Dissipation coefficients for supersymmetric inflatonary models,''
    hep-ph/0603266.

\bibitem{BasteroGil:2010pb} 
  M.~Bastero-Gil, A.~Berera and R.~O.~Ramos,
  ``Dissipation coefficients from scalar and fermion quantum field interactions,''
  JCAP {\bf 1109}, 033 (2011)
  [arXiv:1008.1929 [hep-ph]].
  
\bibitem{BasteroGil:2012cm} 
  M.~Bastero-Gil, A.~Berera, R.~O.~Ramos and J.~G.~Rosa,
  ``General dissipation coefficient in low-temperature warm inflation,''
  JCAP {\bf 1301}, 016 (2013)
  [arXiv:1207.0445 [hep-ph]]. 
  
\bibitem{Ramos:2013nsa} 
  R.~O.~Ramos and L.~A.~da Silva,
  ``Power spectrum for inflation models with quantum and thermal noises,''
  JCAP {\bf 1303}, 032 (2013)
  [arXiv:1302.3544 [astro-ph.CO]].

\bibitem{Kolb:1990vq} 
  E.~W.~Kolb and M.~S.~Turner,
  ``The Early Universe'',
  Front.\ Phys.\  {\bf 69}, 1 (1990).

\bibitem{Berera:2008ar} 
  A.~Berera, I.~G.~Moss and R.~O.~Ramos,
    ``Warm Inflation and its Microphysical Basis,''
      Rept.\ Prog.\ Phys.\  {\bf 72}, 026901 (2009)
      [arXiv:0808.1855 [hep-ph]].


\bibitem{Graham:2009bf} 
  C.~Graham and I.~G.~Moss,
  ``Density fluctuations from warm inflation,''
  JCAP {\bf 0907}, 013 (2009)
  [arXiv:0905.3500 [astro-ph.CO]].


  \bibitem{Mar-Berera-IJMP} 
  M.~Bastero-Gil and A.~Berera,
  ``Warm inflation model building,''
  Int.\ J.\ Mod.\ Phys.\ A {\bf 24}, 2207 (2009)
  [arXiv:0902.0521 [hep-ph]].

\bibitem{Imp} 
  S.~Bartrum, M.~Bastero-Gil, A.~Berera, R.~Cerezo, R.~O.~Ramos and J.~G.~Rosa,
  ``The importance of being warm (during inflation),''
  Phys.\ Lett.\ B {\bf 732}, 116 (2014)
  [arXiv:1307.5868 [hep-ph]].

  \bibitem{BereraGleiserRamos1998}
  A.~Berera, M.~Gleiser and R.~O.~Ramos,
  ``Strong dissipative behavior in quantum field theory,''
  Phys.\ Rev.\ D {\bf 58}, 123508 (1998)
  [hep-ph/9803394].
  
  \bibitem{yokoyamalinde1999}
  J.~Yokoyama and A.~D.~Linde,
  ``Is warm inflation possible?,''
  Phys.\ Rev.\ D {\bf 60}, 083509 (1999)
  [hep-ph/9809409].


 \bibitem{Schwarz:2001vv} 
  D.~J.~Schwarz, C.~A.~Terrero-Escalante and A.~A.~Garcia,
  ``Higher order corrections to primordial spectra from cosmological inflation,''
  Phys.\ Lett.\ B {\bf 517}, 243 (2001)
  [astro-ph/0106020].


\bibitem{Moss:2008yb} 
  I.~G.~Moss and C.~Xiong,
  ``On the consistency of warm inflation,''
  JCAP {\bf 0811}, 023 (2008)
  [arXiv:0808.0261 [astro-ph]].



\bibitem{Hall:2003zp} 
  L.~M.~Hall, I.~G.~Moss and A.~Berera,
  ``Scalar perturbation spectra from warm inflation,''
  Phys.\ Rev.\ D {\bf 69}, 083525 (2004)
  [astro-ph/0305015].

\bibitem{Moss:1985wn} 
  I.~G.~Moss,
    ``Primordial Inflation With Spontaneous Symmetry Breaking,''
      Phys.\ Lett.\  {\bf 154B}, 120 (1985).

\bibitem{BasteroGil:2011xd} 
  M.~Bastero-Gil, A.~Berera and R.~O.~Ramos,
  ``Shear viscous effects on the primordial power spectrum from warm inflation,''
  JCAP {\bf 1107}, 030 (2011)
  [arXiv:1106.0701 [astro-ph.CO]].

\bibitem{Bart} 
  S.~Bartrum, A.~Berera and J.~G.~Rosa,
  ``Gravitino cosmology in supersymmetric warm inflation,''
  Phys.\ Rev.\ D {\bf 86}, 123525 (2012)
  [arXiv:1208.4276 [hep-ph]].


\bibitem{planck2015:params}
  P.~A.~R.~Ade {\it et al.} [Planck Collaboration],
  ``Planck 2015 results. XIII. Cosmological parameters,''
  Astron.\ Astrophys.\  {\bf 594}, A13 (2016)
  [arXiv:1502.01589 [astro-ph.CO]] (Table 4).

\bibitem{camb}
  http://camb.info/

\bibitem{Ade:2013kta} 
  P.~A.~R.~Ade {\it et al.} [Planck Collaboration],
  ``Planck 2013 results. XV. CMB power spectra and likelihood,''
  Astron.\ Astrophys.\  {\bf 571}, A15 (2014)
  [arXiv:1303.5075 [astro-ph.CO]].

\bibitem{Bhattacharya:2005wn} 
  K.~Bhattacharya, S.~Mohanty and R.~Rangarajan,
  ``Temperature of the inflaton and duration of inflation from WMAP data,''
  Phys.\ Rev.\ Lett.\  {\bf 96}, 121302 (2006)
  [hep-ph/0508070].  


\bibitem{CosmoFluids} 
  M.~Bastero-Gil, A.~Berera, I.~G.~Moss and R.~O.~Ramos,
  ``Cosmological fluctuations of a random field and radiation fluid,''
  JCAP {\bf 1405}, 004 (2014)
  [arXiv:1401.1149 [astro-ph.CO]].

\bibitem{planck2015:likelihood}  
 N.~Aghanim {\it et al.} [Planck Collaboration],
  ``Planck 2015 results. XI. CMB power spectra, likelihoods, and robustness of parameters,''
  Astron.\ Astrophys.\  {\bf 594}, A11 (2016)
  [arXiv:1507.02704 [astro-ph.CO]].

\bibitem{Benetti:2016jhf} 
  M.~Benetti and R.~O.~Ramos,
  ``Warm inflation dissipative effects: predictions and constraints from the Planck data,''
  Phys.\ Rev.\ D {\bf 95},  023517 (2017)
  [arXiv:1610.08758 [astro-ph.CO]].

\bibitem{Abazajian:2016yjj} 
  K.~N.~Abazajian {\it et al.} [CMB-S4 Collaboration],
  ``CMB-S4 Science Book, First Edition,''
  arXiv:1610.02743 [astro-ph.CO].
  
\bibitem{Munoz:2016owz} 
  J.~B.~Muñoz, E.~D.~Kovetz, A.~Raccanelli, M.~Kamionkowski and J.~Silk,
  ``Towards a measurement of the spectral runnings,''
  JCAP {\bf 1705}, 032 (2017)
  [arXiv:1611.05883 [astro-ph.CO]].
  
\bibitem{Bastero-Gil:2017wwl} 
  M.~Bastero-Gil, S.~Bhattacharya, K.~Dutta and M.~R.~Gangopadhyay,
  ``Constraining Warm Inflation with CMB data,''
  arXiv:1710.10008 [astro-ph.CO].
 
 
 
%
%
%
%
%
%
%
%
%

\end{thebibliography}


\end{document}